\documentstyle[twocolumn,flushrt,colordvi]{mn}
\newcommand{\br}{\mbox{\bf r}}
\newcommand{\bv}{\mbox{\bf v}}

\newcommand{\hmpc}{\ifmmode\,{\it h }^{-1}\,{\rm Mpc }\else $h^{-1}\,$Mpc\,\fi}
\newcommand{\divv}{{\bf \nabla \cdot v}}
\newcommand{\etal}{{\it et al.}}

\newcommand{\cf}{{\it cf.}}
\newcommand{\eg}{{\it e.g.}}
\newcommand{\ie}{{\it i.e.}}
\newcommand{\kms}{\ifmmode\,{\rm km}\,{\rm s}^{-1}\else km$\,$s$^{-1}$\fi} 
\newcommand{\capt}{\small \baselineskip 12pt}

\title [SEcat vs. PSC$z$]{Consistent $\beta$ values from
density-density and velocity-velocity comparisons}
\author[Zaroubi et al.]{Saleem Zaroubi$^{1}$, Enzo Branchini$^{2}$, 
Yehuda Hoffman$^{3}$, Luiz N. da Costa$^{4}$\\
$^1$ Max Planck Institut f\"{u}r Astrophysik,
Karl-Schwarzschild-Stra{\ss}e 1, 85741 Garching, Germany.\\
$^2$ Dipartimento di Fisica dell'Universit\'a degli Studi ``Roma
 TRE'', Via della Vasca Navale 84, I-00146, Roma, Italy.\\
$^3$Racah Institute of Physics, The Hebrew University,
Jerusalem 91904, Israel.\\
$^4$ European Southern Observatory,
Karl-Schwarschild Strasse 2, 85741, Garching, Germany.\\
}

\begin{document}
%%%%%%%%%%%%%%%%%%%%%%%%%%%%%%%%%%%%%%%%%%%%%%%%%%%%%%%%%%%%%%
\maketitle
\begin{abstract}

We apply a new algorithm, called the Unbiased Minimal Variance
(hereafter UMV) estimator, to reconstruct the cosmic density and
peculiar velocity fields in our local universe from the SEcat catalog
of peculiar velocities comprising both early (ENEAR) and late type
(SFI) galaxies. The reconstructed fields are compared with those
predicted from the IRAS PSC$z$ galaxy redshift survey to constrain the
value of $\beta=\Omega_m^{0.6}/b$, where $\Omega_m$ and $b$ are the
mass density and the bias parameters. The comparison of the density
and velocity fields is carried out within the same methodological
framework, and leads, for the first time, to consistent values of
$\beta$, yielding $\beta=0.57_{-0.13}^{+0.11}$ and $\beta=0.51 \pm
{0.06}$, respectively.

We find that the distribution of the density and velocity residuals,
relative to their respective errors, is consistent with a Gaussian
distribution with $\sigma\approx 1$, indicating that the density field
predicted from the PSC$z$ is an acceptable fit to that deduced from
the peculiar velocities of the SEcat galaxies.

\end{abstract}


%%%%%%%%%%%%%%%%%%%%%%%%%%%%%%%%%%%%%%%%%%%%%%%
\newpage
\section{Introduction}
\label{sec:intro}

In the gravitational instability scenario (\eg, Peebles 1980), mass
density fluctuations and peculiar velocities evolve in an expanding
universe under the effect of gravity. If density fluctuations are
small, linear theory is valid and a simple relation exists between
peculiar velocities, $\bv$, and mass density contrast, $\delta_m$:
\begin{equation} 
\nabla \cdot \bv = -\Omega_m^{0.6} \delta_m,
\label{eq:divv}
\end{equation}
where $\Omega_m$ is the mass density
parameter. Equation~(\ref{eq:divv}) shows
why peculiar motions are so
important in cosmology: they provide a direct probe of the mass
density distribution in the universe.
The mass density fluctuation
field, $\delta_m$, can be deduced from the galaxy observed density
contrast, $\delta_g$, assuming a relation (bias) between the
distribution of galaxies and that of the underlying density. The
simplest relation suggested in the literature is that of linear bias,
namely $\delta_g = b \delta_m$, where $b$ is the linear bias parameter
for a given population of mass tracers. This assumption seems to hold
on very large (linear) scales and it is supported by both
observational evidence (e.g. Baker \etal, 1998 and Seaborne \etal,
1999) and numerical studies (\eg, Blanton \etal, 1999).

Many authors have used galaxies' peculiar velocities and their
redshift space positions to estimate the value of
$\beta=\Omega_m^{0.6}/b$, under the hypotheses of linear theory and
linear biasing. These analyses have been typically carried out using
two alternative strategies. In the so-called density-density
comparisons a 3-D velocity field and a self-consistent mass density
field are derived from observed radial velocities and compared to the
galaxy density field measured from large redshift surveys.  The
typical example is the comparison of the mass density field
reconstructed with the POTENT method (Bertschinger \& Dekel 1989,
Dekel \etal, 1990) from the MARK III catalog of galaxy peculiar
velocities (Willick \etal, 1997a) with the galaxy density field
obtained from the IRAS 1.2 Jy redshift catalog (Sigad \etal, 1998).
The various applications of density-density comparisons to a number of
datasets have persistently led to large estimates of $\beta$,
consistent with unity (see Sigad \etal, 1998 and references
therein). The alternative approach is constituted by the
velocity-velocity analyses. In this second case the observed galaxy
distribution is used to infer a mass density field from which peculiar
velocities are obtained and compared to the observed ones. The
velocity-velocity methods have been applied to most of the velocity
catalogs presently available yielding systematically lower values of
$\beta$, in the range $0.4 - 0.6$ (see Zaroubi 2002a, for a summary of
the most recent results).

Both density-density and velocity-velocity methods have been carefully
tested using mock catalogs extracted from N-body simulations. They
were shown to provide an unbiased estimate of the $\beta$
parameter. Yet, when applied to the same datasets, the discrepancy in
the $\beta$ estimates turned out to be significantly larger than the
expected errors. Accounting for mildly nonlinear motions (e.g. Sigad
\etal, 1998 and Willick \etal, 1996) or allowing for possible deviations
from a pure linear biasing relation consistent with the observational
constraints (see discussion in Somerville \etal, 2001, Branchini \etal
2001) does not explain this discrepancy (Berlind, Narayanan and
Weinberg 2001). Velocity-velocity comparisons are generally regarded
as more reliable as they require manipulation of the denser and more
homogeneous, redshift catalog data. Whereas, the density-density
comparisons involve manipulation of the noisier and sparser velocity
data.  In any case both classes of methods are quite complicated and
it is hard to understand how systematic errors can arise and propagate
through the analysis. Therefore, it is likely that these systematics
affect the $\beta$ parameter estimation.

The purpose of this work is to address, and possibly solve, the
density-density {\it vs.} velocity-velocity dichotomy. We achieve this
goal by using the novel Unbiased Minimal Variance estimator, recently
proposed by Zaroubi (2002b). The UMV estimator allows one to reconstruct an
unbiased cosmological field at any point in space from sparse, noisy
and incomplete data and to 
map it into a dynamically-related cosmic field
($e.g.$ to go from peculiar velocities to overdensities). The UMV is
applied here to the SEcat catalog of peculiar velocities (Zaroubi 2002b)
to reconstruct both the mass density and peculiar velocity fields. These
fields are then compared with the analogous quantities predicted from
the distribution of IRAS PSC$z$ galaxies (Saunders \etal, 2000) 
of density-density and a velocity-velocity analyses. In
Section~\ref{sec:method} we briefly review the basics of the UMV
estimator. The SEcat and PSC$z$ catalogs are presented in
Section~\ref{sec:data}. Error estimation from mock catalogs is
described in \S~\ref{sec:mock}. The density and velocity fields
obtained by applying UMV to SEcat are compared in
Section~\ref{sec:results} with the analogous quantities deduced from
the PSC$z$ dataset. Finally, in Sections~\ref{sec:discussion} we
discuss the results and present our conclusions.

\section{The UMV method}
\label{sec:method}

The derivation and general properties of the UMV estimator have been
already presented 
and discussed by Zaroubi (2002b). 
Therefore, we only review its main properties.
As in the case of Wiener filter (Wiener 1949, Zaroubi \etal, 1995), 
the UMV introduces a
general framework of linear estimation and prediction by minimizing
the variance of the estimator with respect to the underlying signal,
subject to linear constraints on the data. The solution of the
minimization problem yields the UMV estimator, which was shown to be a
very effective tool for reconstructing the large scale structure of the
universe from incomplete, noisy and sparse data (Zaroubi 2002b) .

One of the main drawbacks of the Wiener filter is that it suppresses
the amplitude of the estimated signal.
The suppression factor is roughly equal to the
$Signal^2/(Signal^2 + Noise^2)$ ratio, \ie, in the limit of very poor
signal-to-noise data, which in the context of this work correspond to galaxy 
peculiar velocities, the estimated field approaches zero value.
By contrast, the UMV estimator has been specifically
designed to not alter the values of the reconstructed field at the
locations of the data points, thus avoiding spurious suppression effects.
An unbiased estimate of the reconstructed field at any point in space
is then obtained by interpolating between the data points, according to
the  correlation function assumed {\it a priori}.
Like the Wiener filter the new estimator can be
used for dynamical reconstructions, \ie, to recover one dynamical
field from another measured field, \eg, mass over-density from radial
peculiar velocity. These two properties make the UMV estimator a very
appealing tool for studying the LSS and CMB.

Here we apply the UMV estimator to reconstruct the density and
velocity fields in the local universe from the radial peculiar
velocities of the SEcat galaxies. The data points consist
of a set of observed radial peculiar velocities, $u^o_i$, measured at
positions $\br_i$ with estimated errors $\epsilon_i$,
assumed to be uncorrelated. The observed velocities are thus related to the true
3D underlying velocity field $\bv(\br)$, or to its radial component $u_i$ 
, via

\begin{equation}
u_i^o= \bv(\br_i)\cdot \br_i + \epsilon_i \equiv u_i + \epsilon_i,
\label{vrad}
\end{equation} 

We assume that the peculiar velocity field $\bv(\br)$ and the density
fluctuation field $\delta(\br)$ are related via linear
gravitational-instability theory, $\delta = f(\Omega_m)^{-1} \divv$,
where $f(\Omega_m)\approx\Omega_m^{0.6}$ and $\Omega_m$ is the matter
mean density parameter.  Under the assumption of a specific
theoretical prior for the power spectrum $P(k)$ of the underlying
density field, we can write the UMV estimator of the 3D velocity field
as,
\begin{equation}
\bv^{UMV}(\br)=\langle\bv(\br)u_i\rangle \langle
u_iu_j\rangle^{-1}u_j^o
\label{eq:vumv}
\end{equation} 
and the UMV estimator of the density field as,
\begin{equation}
\delta^{UMV}(\br)=\langle\delta(\br)u_i\rangle \langle
u_iu_j\rangle^{-1}u_j^o.
\label{eq:dumv}
\end{equation} 

Within the framework of linear theory and assuming
that the velocities are drawn from a
Gaussian random field, the two-point velocity-velocity and
density-velocity correlation matrices (bracketed quantities in
eqs.~\ref{eq:vumv} \& ~\ref{eq:dumv}) are readily calculated.  Note
that the normalization of the power spectrum 
drops out of the 
field estimation. 
The calculation of these matrices is discussed elsewhere (G\'orski 1988;
Zaroubi \etal, 1995,1999).

The assumption that linear theory is valid on all scales enables us to
choose the resolution as well. In particular it allows us to use two different
smoothing kernels for the data and for the recovered fields.  In our case no
smoothing was applied to the radial velocity data while we choose to
reconstruct the density and velocity fields with a finite Gaussian
smoothing of radius, $R$. This choice alters the velocity-velocity and
density-velocity correlation functions that appear in the first
bracketed terms of the right hand side of eqs.~\ref{eq:vumv} and
~\ref{eq:dumv}, respectively, by introducing a multiplicative term
$\exp[-k^2 R^2/2] $ in the model power spectrum. 

The amplitude of the reconstructed matter density field given in
equation~\ref{eq:dumv} is proportional to $f(\Omega_m)^{-1}$, while
that of the density field obtained the PSC$z$ galaxy distribution is
proportional to the biasing parameter, $b$. Therefore, the comparison
between these density fields will constrain the value of $\beta =
f(\Omega_m)/b$. The velocity field reconstruction from the SEcat data
set, however, is independent of $\Omega_m$ and $b$. Hence in the
velocity-velocity comparison $\beta$ enters as a solution of
eq.~\ref{eq:divv} for the PSC$z$ velocities, with the matter density
given as the ratio $\delta({\em PSCz})/b$, where $\delta({\em PSCz})$
is the PSC$z$ density field.

The error covariance matrix, or variance of the residuals, of the
reconstructed density and velocity fields could be 
calculated theoretically (Zaroubi 2002b). However, in order to give a more complete
account of the various errors that enter the calculation (\eg,
nonlinear effects) here we choose to calculate the error from mock
catalogs (see section~\ref{sec:mock}).

It is interesting to compare the UMV algorithm with other methods of
reconstruction used for similar purposes. In the context of
mass-density reconstruction from radial peculiar velocities,
comparison with the POTENT algorithm (Bertschinger \& Dekel 1989,
Dekel \etal, 1990) is of special interest. The main assumption behind
the POTENT algorithm is that the flow field, smoothed on large scale,
is derived from a velocity potential. The potential flow assumption is
a direct result of linear theory but can also be employed in the
quasi-linear regime until the onset of shell crossing, when an
extension of eq.~\ref{eq:divv} applies (Nusser \etal, 1991). POTENT
could be viewed as a direct inversion method which uses the minimum
amount of assumptions, but suffers from the problems of direct
deconvolution of very noisy data.

In conclusion, the UMV reconstruction can be regarded as a compromise
between the POTENT algorithm, which assumes no regularization but
might be unstable to the inversion problem of deconvolving highly
noisy data, and the WF algorithm, which takes into account the
correlation between the data points and therefore stabilizes the
inversion, but constitutes a biased estimator of the underlying field.

\section{The datasets}
\label{sec:data}

The main dataset used here is the SEcat catalog of galaxy peculiar
velocities which results from the merging of 1300 spiral galaxies
taken from the SFI catalog (Giovanelli \etal, 1997a and 1997b and
Haynes \etal, 1999a and 1999b) and about 2000 early type galaxies from
the ENEAR catalog grouped into $\approx$ 750 independent objects (da
Costa \etal, 2000). For each object the radial velocity and inferred
distance, corrected for Malmquist bias (Freudling \etal, 1995, da Costa
\etal, 2000), are provided along with the velocity errors that typically
amount to $\sim 19\%$ of the galaxy distance.  

\begin{figure*}
\setlength{\unitlength}{1cm} \centering
\begin{picture}(18.,25.)
\put(-1.8, 16.){\includegraphics{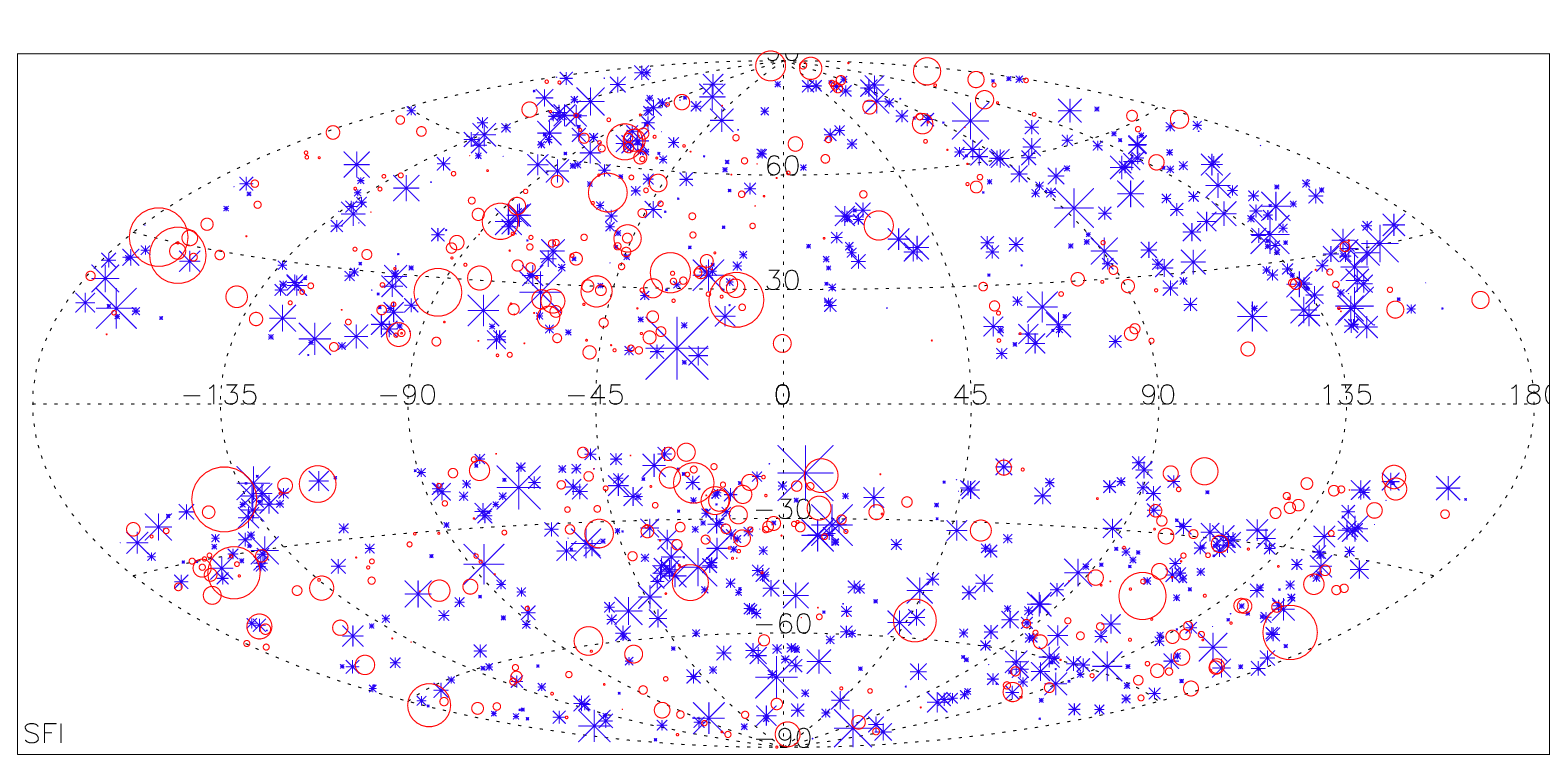}}
\put(7.4,16.){\includegraphics{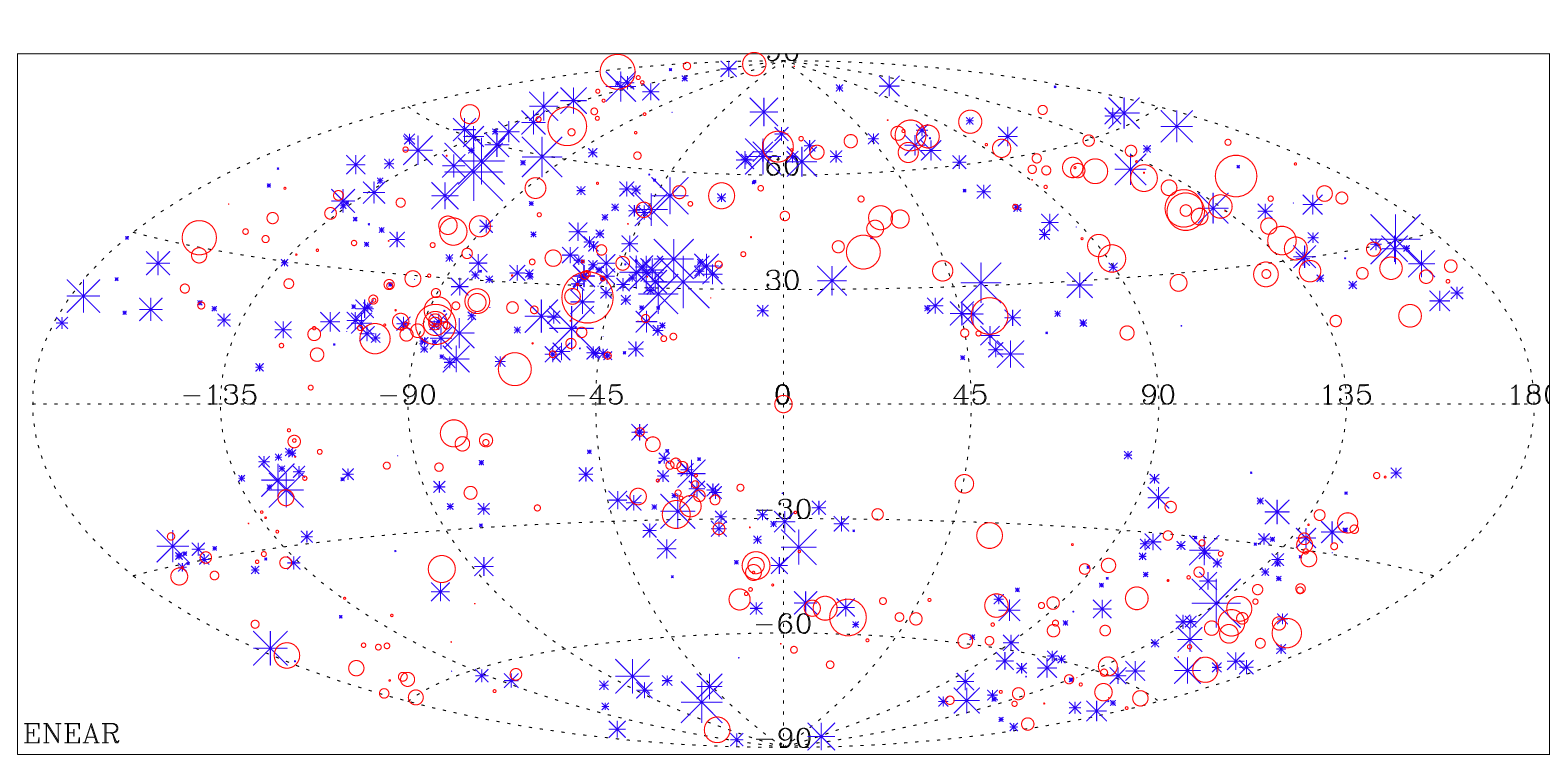}}
\put(-3, 4.5){\includegraphics{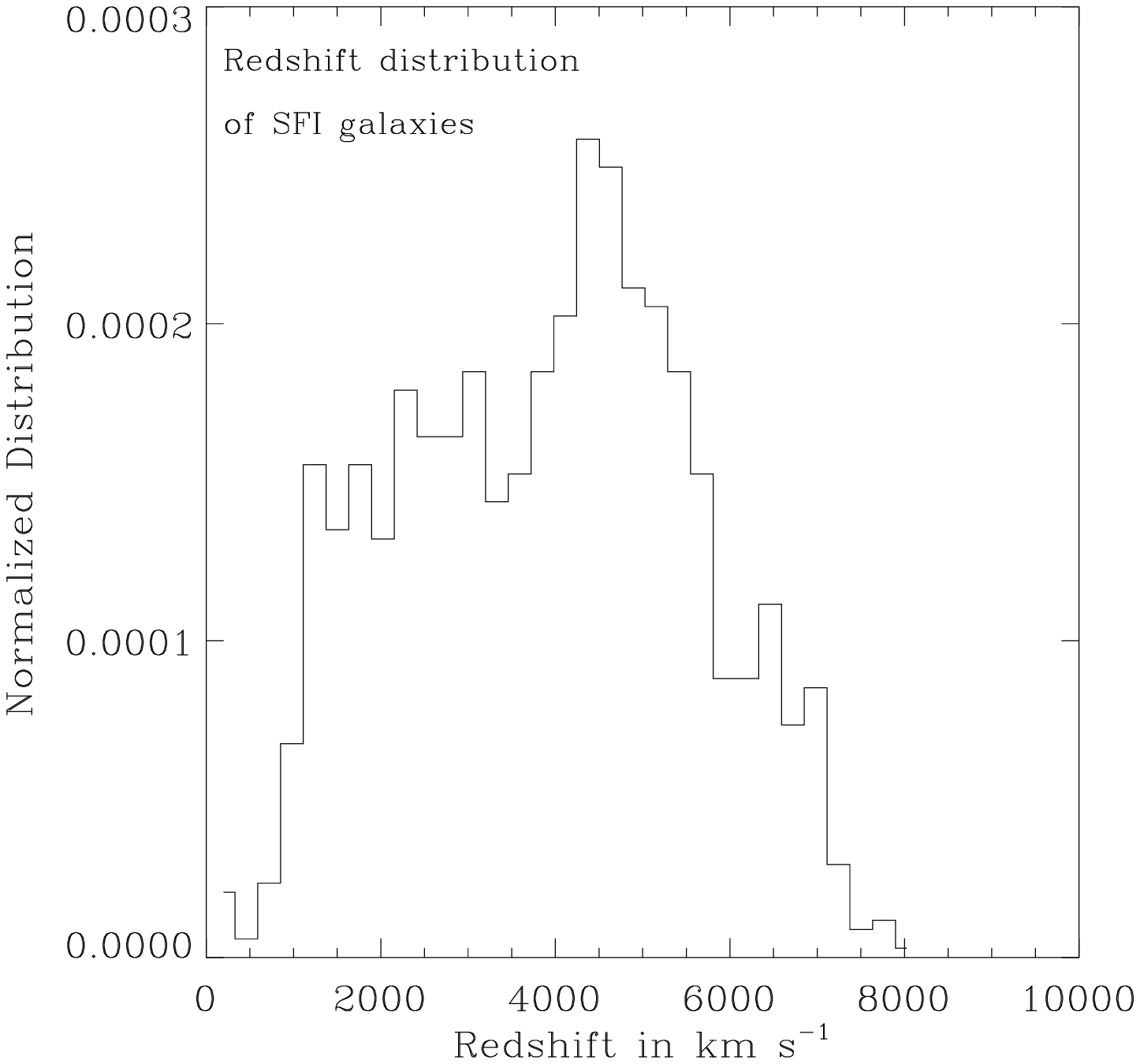}}
\put(6.2, 4.5){\includegraphics{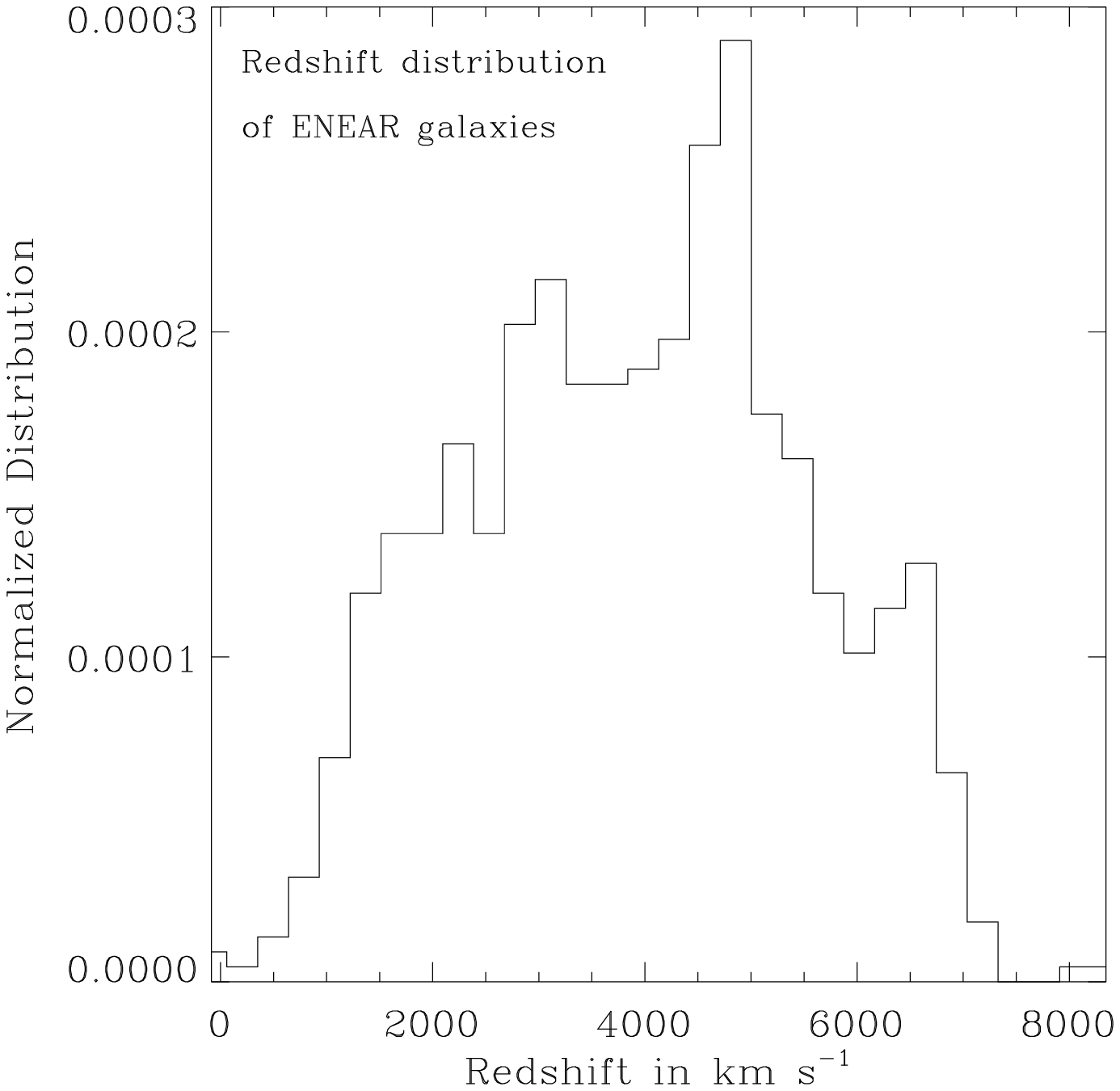}}
\put(-3, -3.8){\includegraphics{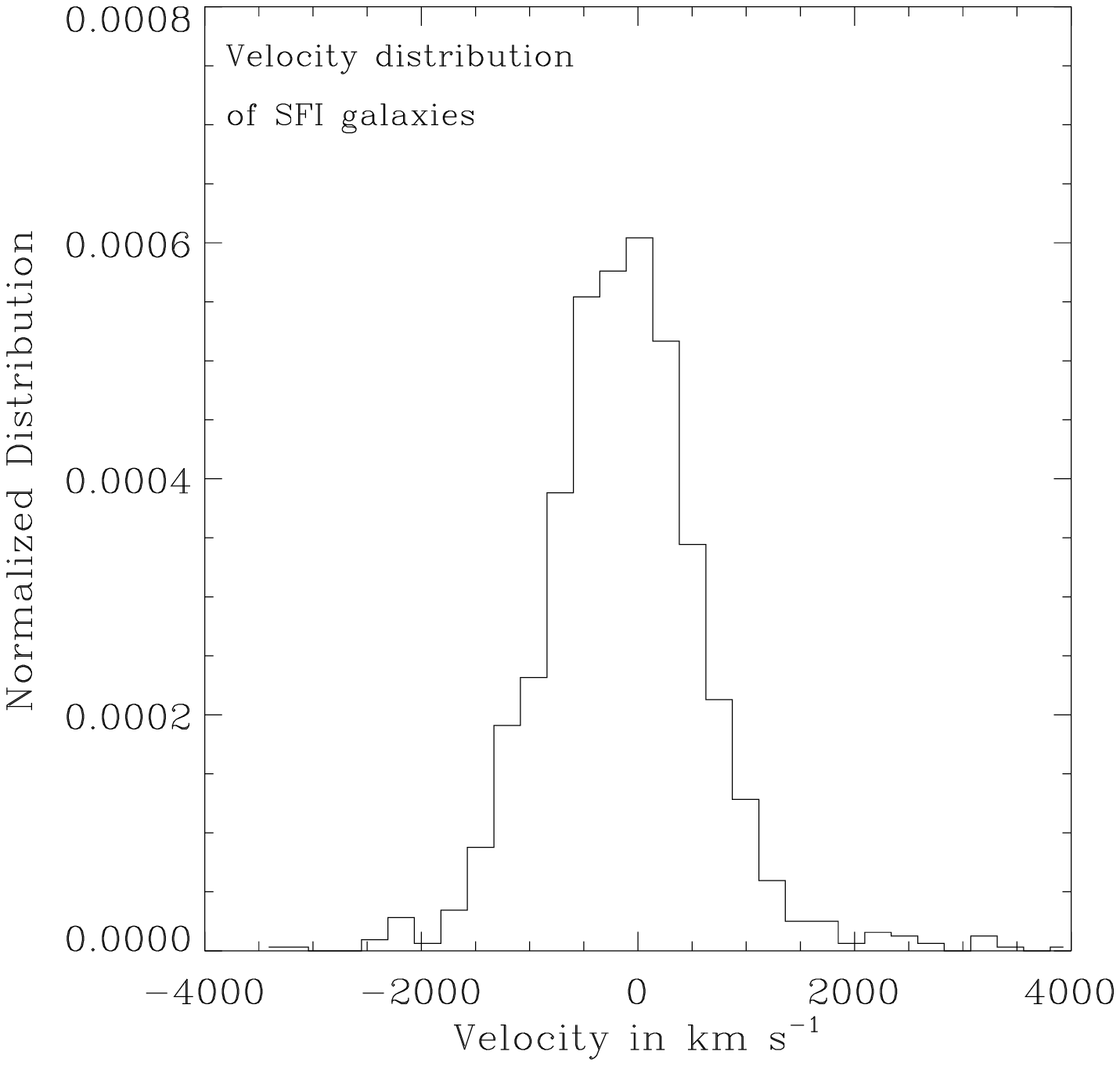}}
\put(6.2, -3.8){\includegraphics{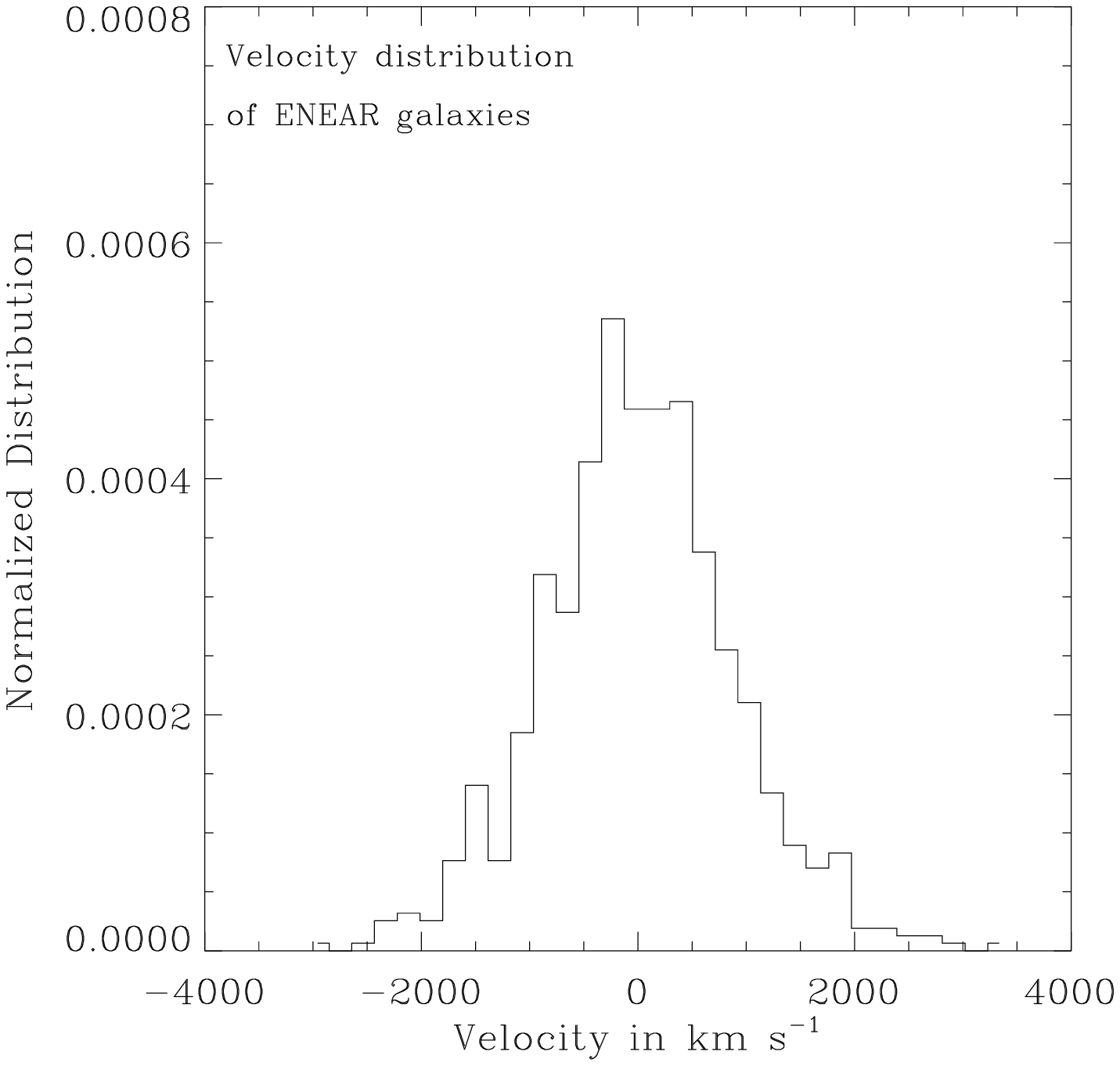}}
\end{picture}
\vskip -4 truecm
\caption{\capt Upper panels: The projected distribution of the SFI
(left panel) and ENEAR (right panel) galaxies in galactic coordinates.
Crosses indicate positive peculiar velocities, open circles negative;
and the size of the symbols is proportional to the amplitude of  their
peculiar velocity.
Middle panels: The redshift distribution of the SFI (left panel) and
the ENEAR (right panel) galaxies. Lower panels: The peculiar velocity
distribution of the the SFI (left panel) and the ENEAR (right panel)
galaxies.
}
\label{fig:datasets}
\end{figure*}    
Merging different peculiar velocity catalogs may result in spurious
flows and lead to systematic errors in the reconstruction
procedure. However, several pieces of evidence indicate that this is
not a serious problem for the SEcat catalog.  First of all, both the
SFI and ENEAR catalogs are intrinsically homogeneous as they consist
of uniform data covering most of the sky.  Secondly, as shown in
Bernardi \etal (2002b), the distance relations independently
calibrated in the two sets are consistent with each other and the
difference in the Hubble constant deduced from each catalog is
$\approx 5 \pm 10$ \kms/Mpc. Finally, different statistical analyses
carried out using either samples lead to consistent results (\eg,
Borgani \etal, 2001, da Costa \etal, 2001, Nusser \etal, 2001, Zaroubi
\etal, 2001 and references therein). Some of the main characteristics
of these samples are summarized in Figure ~\ref{fig:datasets}. The sky
distribution (upper panels), redshifts (central panels) and peculiar
velocities (lower panels) of the SFI and ENEAR galaxies are quite
similar, especially when accounting for the expected tighter spatial
correlation and higher peculiar velocities of the ENEAR early type
galaxies.  Further evidence for the consistency of the two catalogs
will be provided in section~\ref{sec:results} (\eg, see
Figures~\ref{fig:density_compare} and~\ref{fig:velocity_compare}) .

The substantial number of galaxies and the large sky coverage (the
unobserved region is given by a zone of avoidance of about
$15^{\circ}$ about the Galactic-plane) allow a dense and uniform
sampling of the velocity field. Moreover, since SEcat contains both
early and late type galaxies, we can sample both high and low density
regions and therefore minimize possible biases that might have affected
other analyses based on a single population of objects.

To estimate the $\beta$ parameter one needs to compare the densities
and/or velocities reconstructed from a radial velocity survey with those
recovered from a direct probe of the density field, namely a redshift
survey. Here we use the models obtained by Branchini \etal (1999) from
the distribution of IRAS PSC$z$ galaxies under the assumptions of
linear biasing and linear theory. The PSC$z$ redshift survey (Saunders
\etal, 2000) provides the angular positions and redshifts of $\sim
15000$ IRAS galaxies distributed over almost all sky (the zone of
avoidance is about $8^{\circ}$) with a median redshift of 8500 \kms
and is therefore suitable for modeling the density and velocity fields
within the same region in which the UMV reconstruction is performed.
The typical error associated with the PSC$z$ density and velocity models
are significantly smaller than those reconstructed from the SEcat
velocities and therefore will be ignored in the subsequent analysis.

\section{Mock catalogs and error estimates}
\label{sec:mock}

\begin{figure*}
\setlength{\unitlength}{1cm} \centering
\begin{picture}(15.,18.)
\put(-3,-3.){\includegraphics{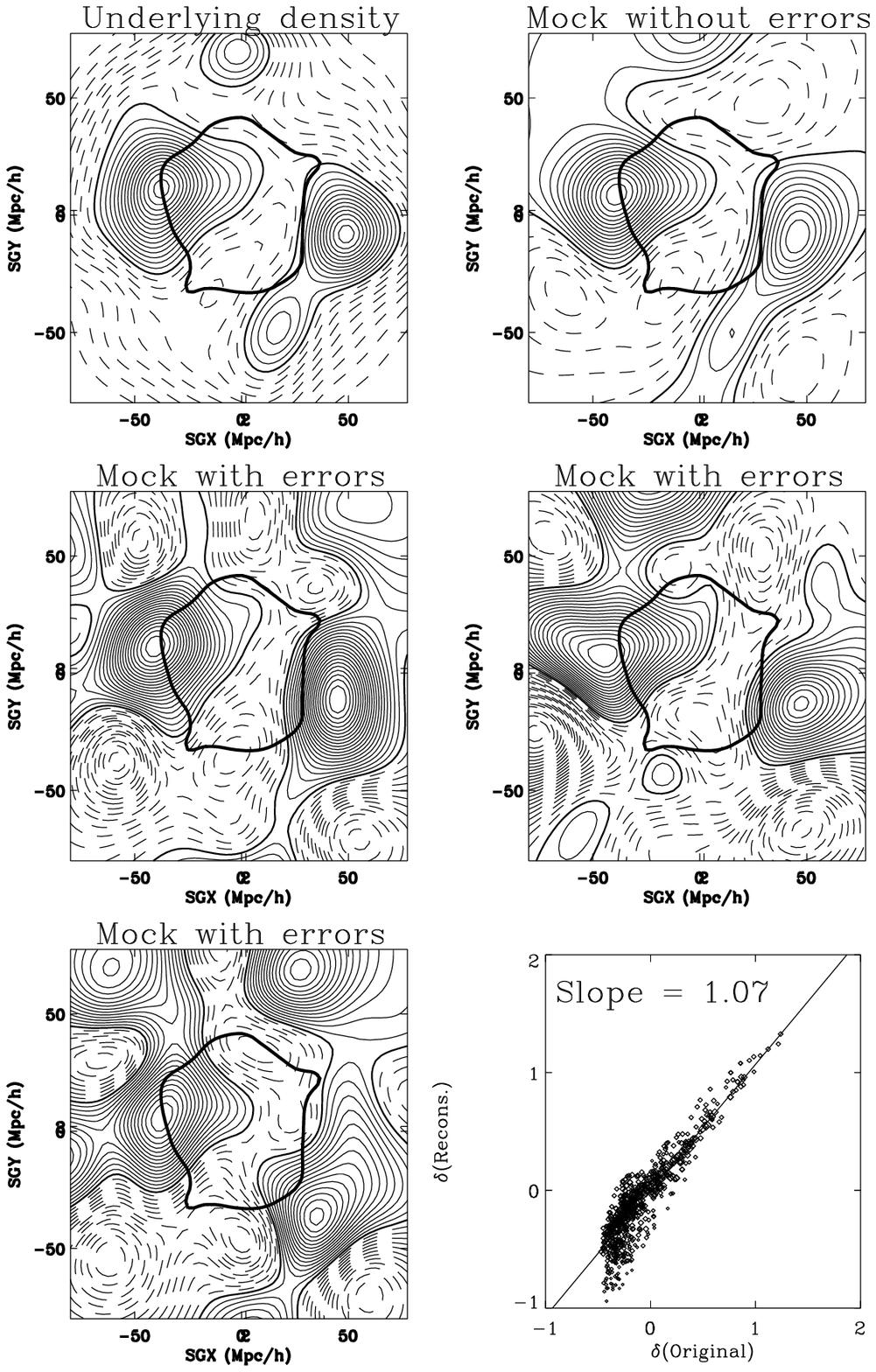}}
\end{picture}
\vskip -.5 truecm
\caption{\capt Comparison of the original and reconstructed G12
smoothed density maps on the mock Supergalactic plane.  In all panels
the solid and dashed line contours denote positive and negative
densities respectively. The bold-solid line denotes the zero level
density.  Contour spacing is 0.1. The very thick solid contour marks
the area within which the reconstruction errors are less than 0.2. The
upper left hand panel shows the underlying density field of the
simulation. The degradation of the original density map towards the
edges is spurious and due to the finite size of the smoothing radius. The
upper right hand panel shows the reconstructed density from the SEcat
mock catalog before adding errors to the distances and velocities. The
other three maps show reconstruction from SEcat mock catalogs with
realistic noise. The lower right-hand panel shows a typical scatter
plot of the density within the comparison region of the original
vs. the reconstructed density from one of the mock catalogs (with
noise); the points used in the scatter plot are 1/10 randomly sampled
from the grid points with estimated reconstruction errors
less than 0.2. }
\label{fig:density_mock}
\end{figure*}    

\begin{figure*}
\setlength{\unitlength}{1cm} \centering
\begin{picture}(18.,9.)
\put(-3,-3.5){\includegraphics{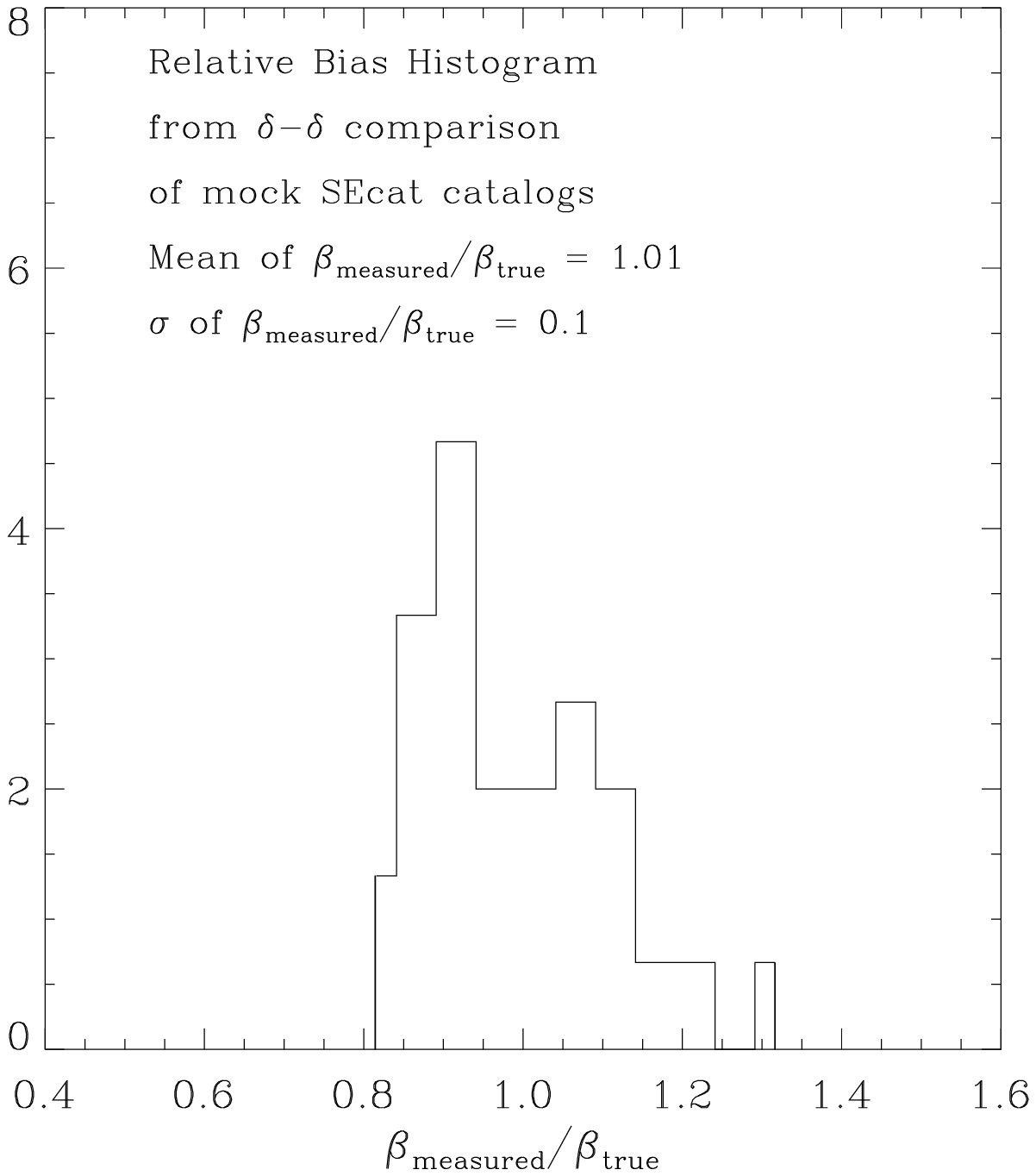}}
\put(5.7,-3.5){\includegraphics{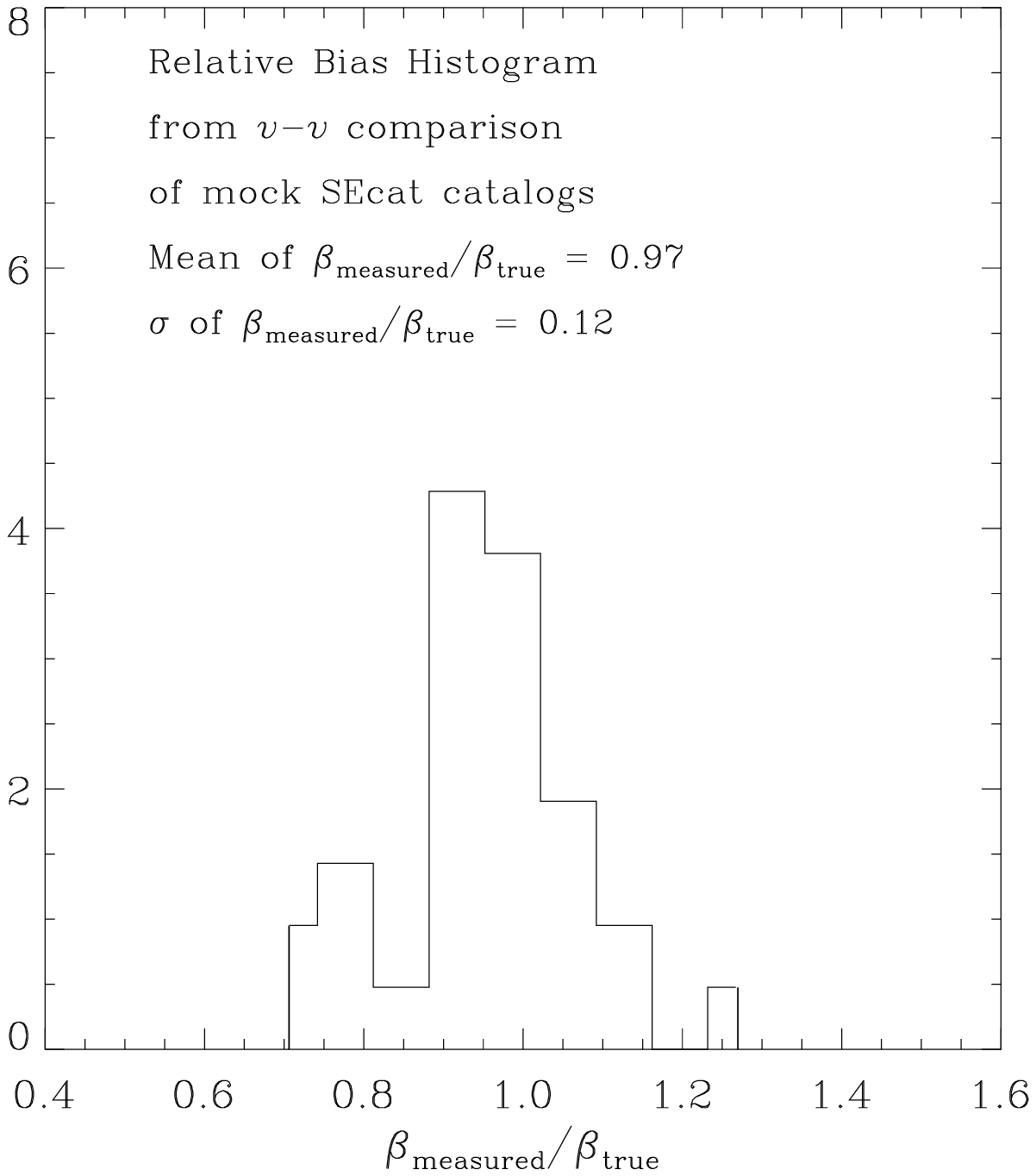}}
\end{picture}
\vskip -.5 truecm
\caption{\capt The distribution of the value of $\beta$ relative to
the real one as estimated from 100 mock SEcat catalogs in the
density-density comparison (left panel) and velocity-velocity
comparison (right panel). The value of the measured $\beta$ for the
catalog without noise is $\beta=0.99$ and $1.01$ for the density and
velocity reconstruction respectively.
}
\label{fig:betafrommock}
\end{figure*}    
To test the performance of the method when applied to the SEcat
catalog, we construct mock catalogs based on the ``Constrained
Realization GIF'' simulation carried out by Mathis \etal\ (2001).
This simulation starts from initial conditions with a smoothed linear
density field which matches that derived from the IRAS 1.2 Jy galaxy
survey and tracks the formation and evolution of all dark matter halos
more massive than $10^{11}$ solar masses out to a distance of $8000$
\kms from the Milky Way up to the current epoch.  Galaxies in the
original mock catalog are sampled from the N-body simulation at the
same position of the real galaxies. Realistic mock catalogs
are obtained by assigning errors to the position and velocity of
mock galaxies consistent with observations. We construct 100 mock
catalogs that differ only in the realization of errors added to the
position and velocity of the galaxies.  Measured peculiar velocities
in the mock catalogs are then obtained after performing a Malmquist
bias correction according to the recipe given by Willick \etal\
(1997a).

The density and 3D velocity fields within a region of 60 \hmpc,
smoothed with a Gaussian filter of radius of 12 \hmpc (G12
hereafter), were reconstructed from each of the 100 mock catalogs.
Both N-body and reconstructed density and velocity fields were
specified on a regular grid with a mesh size of $2.5 \hmpc$ and
reconstructed radial velocities at the actual location of the data
points were interpolated from the grid.  Errors in the reconstructed
densities and peculiar radial velocities were estimated from the 100
Monte Carlo realizations.

Note that we do not use the peculiar velocities of the mock galaxy
catalogs of Mathis \etal\ (2001). Instead, the peculiar velocities of
our mock galaxies are obtained from those of of the dark matter
particles.  This implies neglecting the so called velocity bias which
is expected to be small on the large smoothing scale involved in our
reconstruction (see \eg, Carlberg 1994).

Figure~\ref{fig:density_mock} shows the quality of the reconstruction
from mock SEcat catalogs. The comparison of the original underlying
density with the one reconstructed from ideal mock catalogs (i.e. with
no velocity errors) in a region in which the estimated reconstruction
errors are $\le 0.2$ shows that their relation is well described by a
linear function with a slope of $0.99$ (see below). This demonstrates
that the sampling density of the data is sufficiently high. Note that
the density of the original catalogs close to the boundaries is
suppressed. This effect is due to finite smoothing length and does not
affect our analyses which are carried out to distances well within the
edges of the mock catalogs.  The other three maps shown in Figure 2
were randomly chosen from the 100 mock catalogs reconstructions. The
similarity of the reconstructions are quite evident, especially within
the region of small reconstruction errors.  The scatter plot
quantifies this similarity for one of the mock catalogs. An inspection
of the radial peculiar velocity reconstruction shows a similar quality
of results.

Left panel of Figure~\ref{fig:betafrommock} shows $\beta$ as estimated
by applying the UMV reconstruction algorithm to the mock catalogs and
comparing the results with the true density and velocity fields using
the following $\chi^2$ statistic,
\begin{equation}
\chi^2 = {1 \over N}\sum_{\sigma_\delta \le 0.2} {\left[\delta_i({\rm Mock}) - \beta (\delta_i({\rm Original}) +
\Delta\delta)\right]^2\over \sigma_\delta^2},
\label{eq:chi2_mock}
\end{equation}
where $\delta_i({\rm Mock})$ is the density as reconstructed
from the mock SEcat catalog; $\delta_i({\rm Original})$ is the original
density, and $\sigma_\delta$ is the 
% E correct ?
% standard deviation at each of the
rms difference between the reconstructed densities from the mock
catalogs and the original density at the same point in space.  The
free parameters here are $\beta$ and the offset between the two
fields, $\Delta\delta$. The latter is introduced to account for the
uncertainty in defining the mean density of the sample.  This offset
is expected to be zero in the mock catalog analyses and can only be
non trivial when comparing the density fields obtained from two
different catalogs (such as the true SEcat and PSC$z$ catalogs).  The
density-density comparison is carried out over $N$ points, randomly
selected from the $10\times N$ at which the estimated errors are less
than 0.2.  The expected values of $\beta$ and $\Delta\delta$ are
unity~\footnote[1]{The density reconstruction is performed hereafter
normalizing with the correct value of $f(\Omega_m=0.3)$ therefore the
comparison is expected to yield a bias parameter of unity.} and zero
respectively. The left panel of Figure~\ref{fig:betafrommock} shows
that the estimated $\beta$ is unbiased and has a mean and variance of
$1.01$ and $0.1$, respectively. The estimated value of $\Delta\delta$
is $-0.03$ and has a variance of $0.04$.

Naturally, most of the grid points used in the comparison -- even
after diluting their number by a factor of 10 -- do not have
statistically independent errors. Therefore, the statistic used in
eq.~\ref{eq:chi2_mock} is not optimal.  In other words, had the
data points used in eq.~\ref{eq:chi2_mock} been independent then it
would indeed be sufficient to use the likelihood contours obtained
after minimizing the $\chi^2$ statistic to determine the
uncertainty of the results. On the other hand, had these data point
been totally dependent 
with
about only one degree-of-freedom then the error obtained from the likelihood analysis
would be a gross underestimate unless the {\it
cosmic-variance like} errors over the whole sample were also 
estimated and taken into account.
Normally, the best
way to go about estimating the scatter in the results that takes into
account the partial dependency of the data points and their limited
number is to perform a very large number of Monte Carlo simulations
-- of the order of $10^2$-$10^3$ -- to ensure an accuracy of a few
percent. This however is very time consuming and not feasible with
available computer resources. Therefore, we choose to assume that
the errors are totally independent and use eq.~\ref{eq:chi2_mock} in
order to estimate the most likely $\beta$ and $\Delta\delta$
values. The uncertainties are then determined by adding in quadrature 
the likelihood errors and the uncertainties obtained from the scatter in
the value of $\beta$, obtained from the 100 mock catalogs. In our
sample, these two sources of errors are not totally independent.
However, by treating them as such allows one to obtain a conservative upper
limit on the errors.

\begin{figure}
\setlength{\unitlength}{1cm} \centering
\begin{picture}(8.,9.)
\put(-2,-7.5){\includegraphics{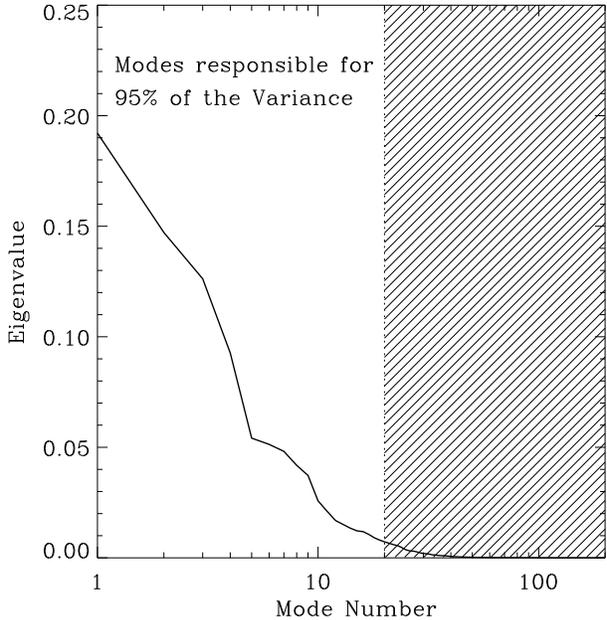}}
\end{picture}
\vskip -.5 truecm
\caption{\capt The sorted spectrum of the eigenvalues of the noise
correlation matrix. The modes that account for $95\%$ of the variance
are the first 20.  }
\label{fig:eigenvalues}
\end{figure}    
In order to validate our approach it is important to estimate the
number of degrees-of-freedom in the sample, ${\cal N}_{dof}$.  In the case
of no noise correlations except those introduced by the G12 smoothing,
${\cal N}_{dof} \approx 23 $ in a sphere of radius $50$\hmpc, which
represents the ratio of the total volume to the effective volume of
the G12 filter. In the case at hand, however, the calculation is more
subtle and involves numerical estimation of the noise correlation
matrix from the 100 Monte-Carlo simulations at each grid point with
error less than 0.2 and finding its eigenvalues. Then ${\cal N}_{dof}$
is identified with the number of significant eigenvalues of this
matrix (see Zaroubi \etal, 1995 for the treatment of a similar
problem).  Specifically, ${\cal N}_{dof}$ is the number of the highest
eigenvalues that account for $95\%$ of the variance, found here to be
about $20$ (see Figure~\ref{fig:eigenvalues}). The eigenvalues of most
of the remaining eigenmodes drop by orders of magnitude. This number,
$20$, reflects the number of effective smoothing volumes and the
additional correlation introduced by the UMV filter.

The same strategy is adopted for the velocity-velocity comparison.
In this case the statistic of choice  is,
\begin{equation}
\chi^2 = {1\over N_{data}} \sum_{Data Points} {\left(u_i({\rm Mock}) -
\beta u_i({\rm Original}) - \Delta H_{\circ}r_i\right)^2\over
\sigma_v^2},
\label{eq:chi2_vvmock}
\end{equation}
where $u({\rm Mock})$ and $u({\rm Original})$ are the radial
velocities of the mock and original data respectively. $N_{data}$ is
the number of data points, $\Delta H_{\circ}$ is the offset in Hubble
constant, $r_i$ is the distance of the data point $i$ and $\sigma_v$
is the uncertainty in the radial velocity as obtained from the mock
catalogs.  The velocity-velocity comparison is carried out over the
radial velocities at the location of the data points. The argument
regarding the $\chi^2$ statistic used here is identical to the one
discussed for the density-density comparison. Right panel of
Figure~\ref{fig:betafrommock} shows that the mean and variance of
$\beta$ in this case are $0.97$ and $0.12$, respectively. The offset
in Hubble constant is $0.2 \pm 0.5$\kms/Mpc. The value of $\beta$
estimated from the noise-free mock catalog is $1.03$.  From the noise
correlation matrix we estimate ${\cal N}_{dof} \approx 17$.  The lower
value obtained here reflects the longer range of velocity correlation.

\begin{figure*}
\setlength{\unitlength}{1cm} \centering
\begin{picture}(18.,17.)
\put(-4,-.5){\includegraphics{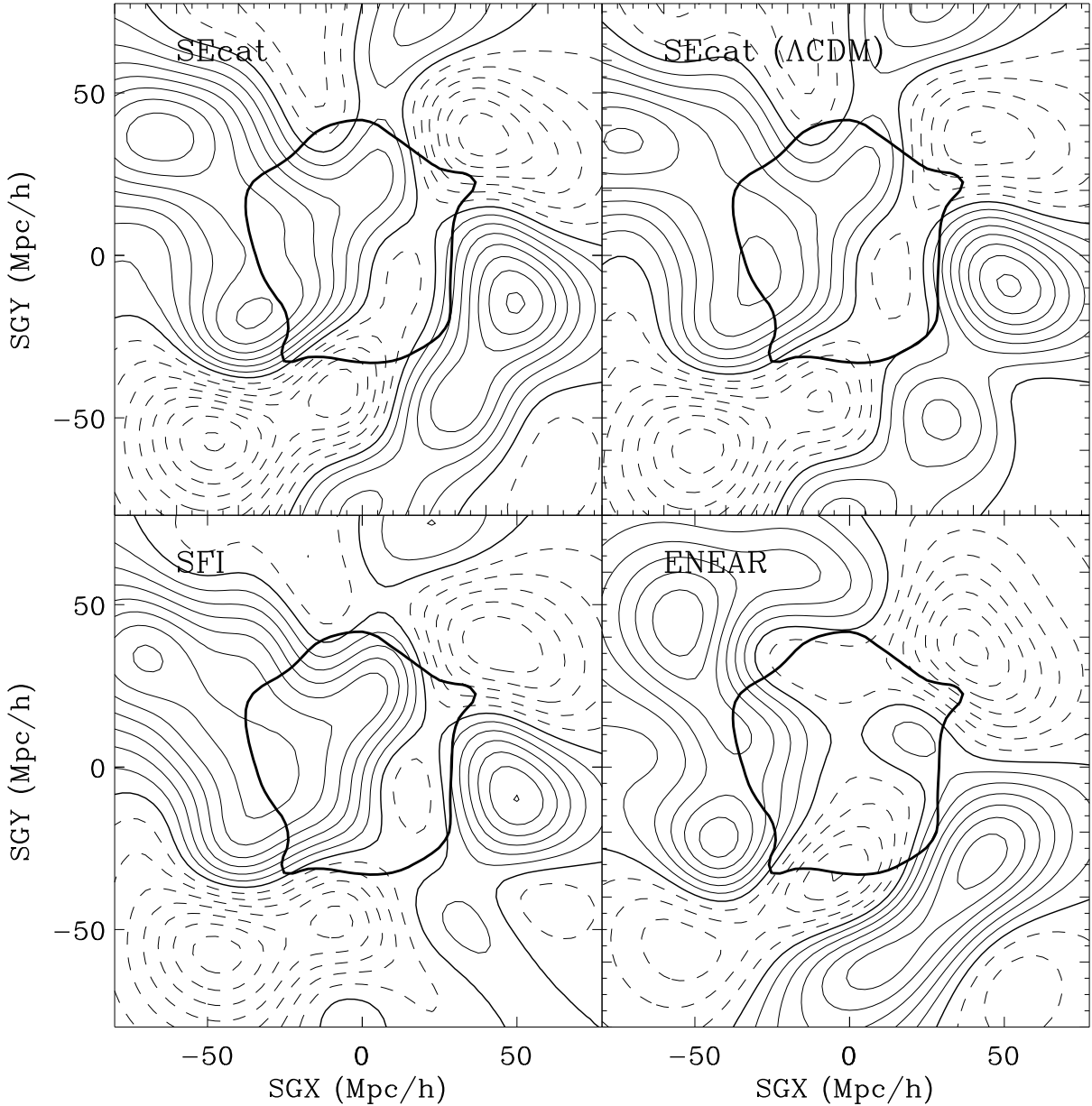}}
\end{picture}
\vskip -.5 truecm
\caption{\capt The G12-smoothed, overdensity field
on the Supergalactic plane reconstructed from SEcat, ENEAR and SFI catalogs. 
All reconstructions
assume a Standard CDM power spectrum apart from the map shown in the upper
right panel for which we have assumed an
$\Omega_m=0.3$ $\Lambda$CDM power spectrum. The very thick solid
contour marks the area within which the reconstruction errors from the
SEcat catalog are less than 0.2. In all panels the solid and
dashed line contours denote positive and negative densities
respectively. The bold-solid line denote the zero level density.
Contour spacing is 0.1. 
}
\label{fig:density_compare}
\end{figure*}    

Willick and Strauss (1997b) have suggested in their VELMOD method to
estimate $\beta$ from velocity data by minimizing residuals and their
correlations simultanously with minimizing the likelihood, therefore,
allowing the calibration of the distance estimator while evaluating
$\beta$. Whereas in our analysis we assume that the distance indicator
has been calibrated independently.

The analysis of the mock catalogs shows that the UMV returns an
unbiased estimate of the underlying density and velocity fields from
which $\beta$ can be determined with an accuracy of $10\%$.  This
error is purely random and takes into account the scatter of the best
fit $\beta$ value but does not include the scatter in the estimation
of each individual value of $\beta$ which will be added later.

This analysis could in principle be extended to smaller smoothing
kernels. Besides the validity of linear theory required in this
analysis, the density of the sky coverage might be also an important
factor since an insufficient coverage might result in the dominance if
noise in most of the sampling volume.  An extension of the analysis to
a $900 \kms$ smoothing kernel (hereafter, G9) shows that the results of
the analysis are consistent with the G12 comparison but the
uncertainties are much higher especially in the density-density
comparison. Therefore, the smoothing kernel used in the rest of the
paper is G12.

\section{The field-field comparison}
\label{sec:results}

\begin{figure}
\setlength{\unitlength}{1cm} \centering
\begin{picture}(8.,9.)
\put(-2.,-7.5){\includegraphics{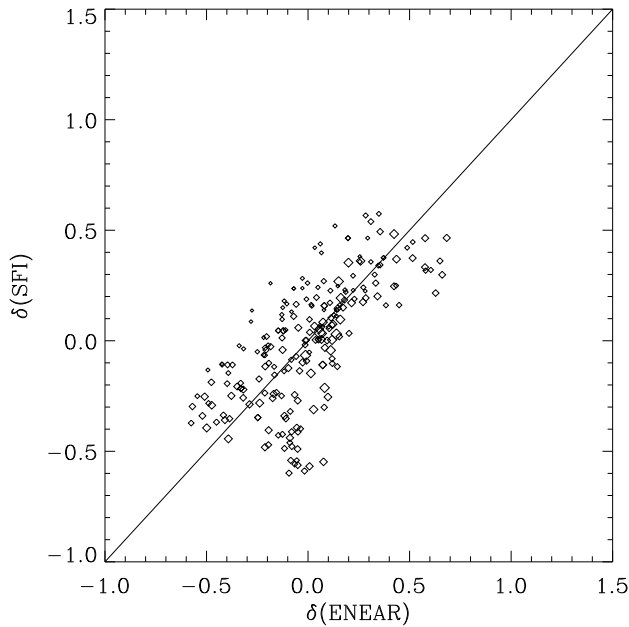}}
\end{picture}
\vskip -.5 truecm
\caption{\capt A quantitative comparison between the SFI and ENEAR
G12-smoothed. reconstructed densities.  The densities of both SFI and
ENEAR were reconstructed on a grid with mesh size of 2.5 \hmpc\. The
densities shown in the scatter plot are from gridpoints randomly
selected with a rate 1/10 from those with reconstruction errors less
than 0.2. The size of the symbols is inversely proportional to their
errors. The solid line with a slope of unity has been drawn to guide
the eye.  The agreement between the two reconstructions is very good
except for a small number of points with $\delta(\mathrm{SFI}) \approx
-0.5$ and $\delta(\mathrm{ENEAR}) \approx 0$.  }
\label{fig:enear_vs_sfi}
\end{figure}    

Here we apply the UMV estimator to the true SEcat catalog to obtain
the G12-smoothed density and velocity fields assuming a power spectrum
of a flat CDM model with $h=0.5$, $\Omega_m=1.0$ as a prior. These
cosmological parameters determine the shape of the power spectrum only
and are not used anywhere else in the analysis. Unlike the outcome of
the Wiener Filter estimator, the UMV-reconstructed density and
velocity fields, for dense sampling of the sky, are unbiased and
therefore can be used for quantitative comparisons.

\begin{figure*}
\setlength{\unitlength}{1cm} \centering
\begin{picture}(18.,20.)
\put(-2., 4.){\includegraphics{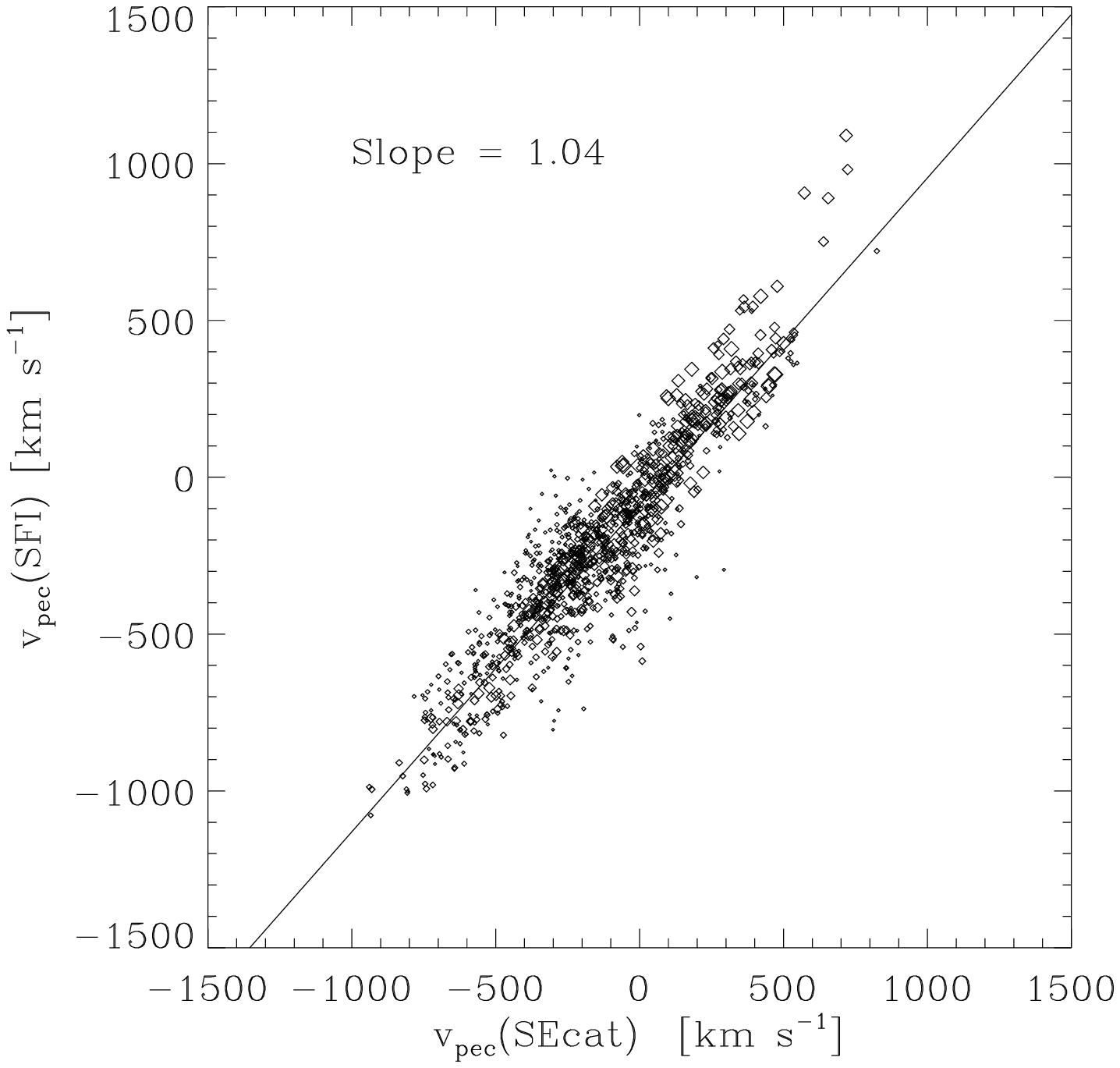}}
\put(7.,4.){\includegraphics{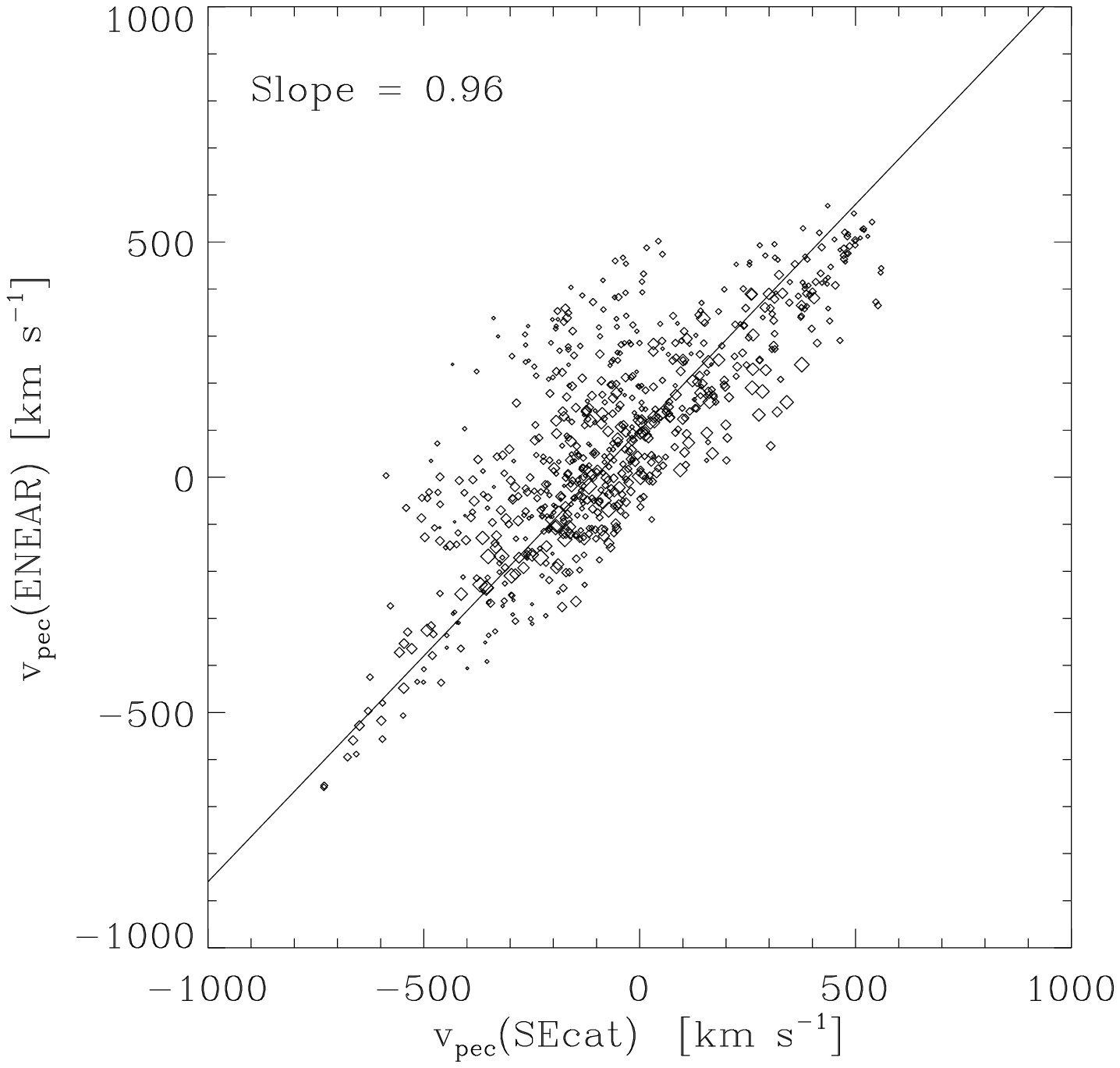}}
\put(-2., -4.5){\includegraphics{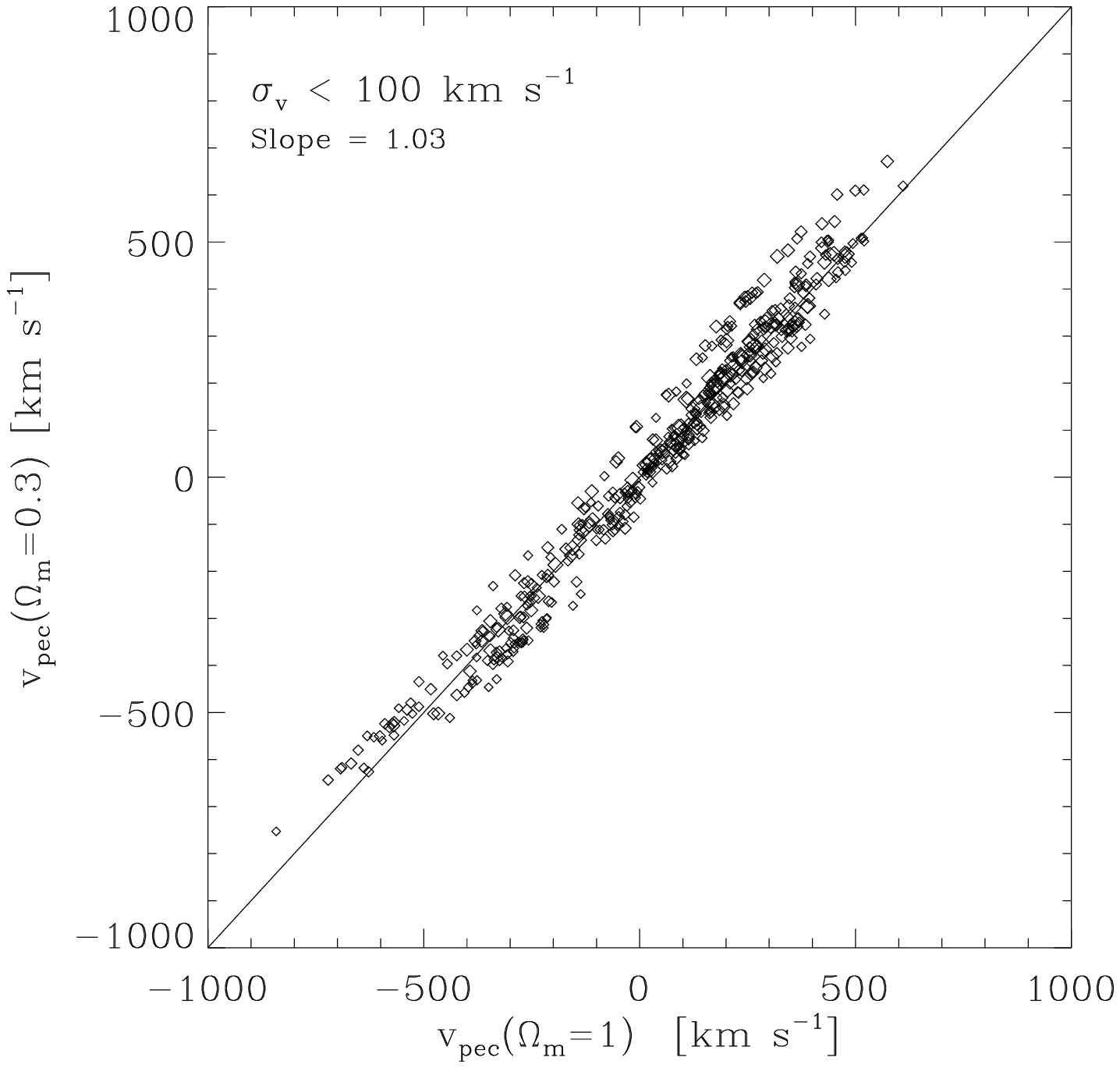}}
\put(7.,-4.5){\includegraphics{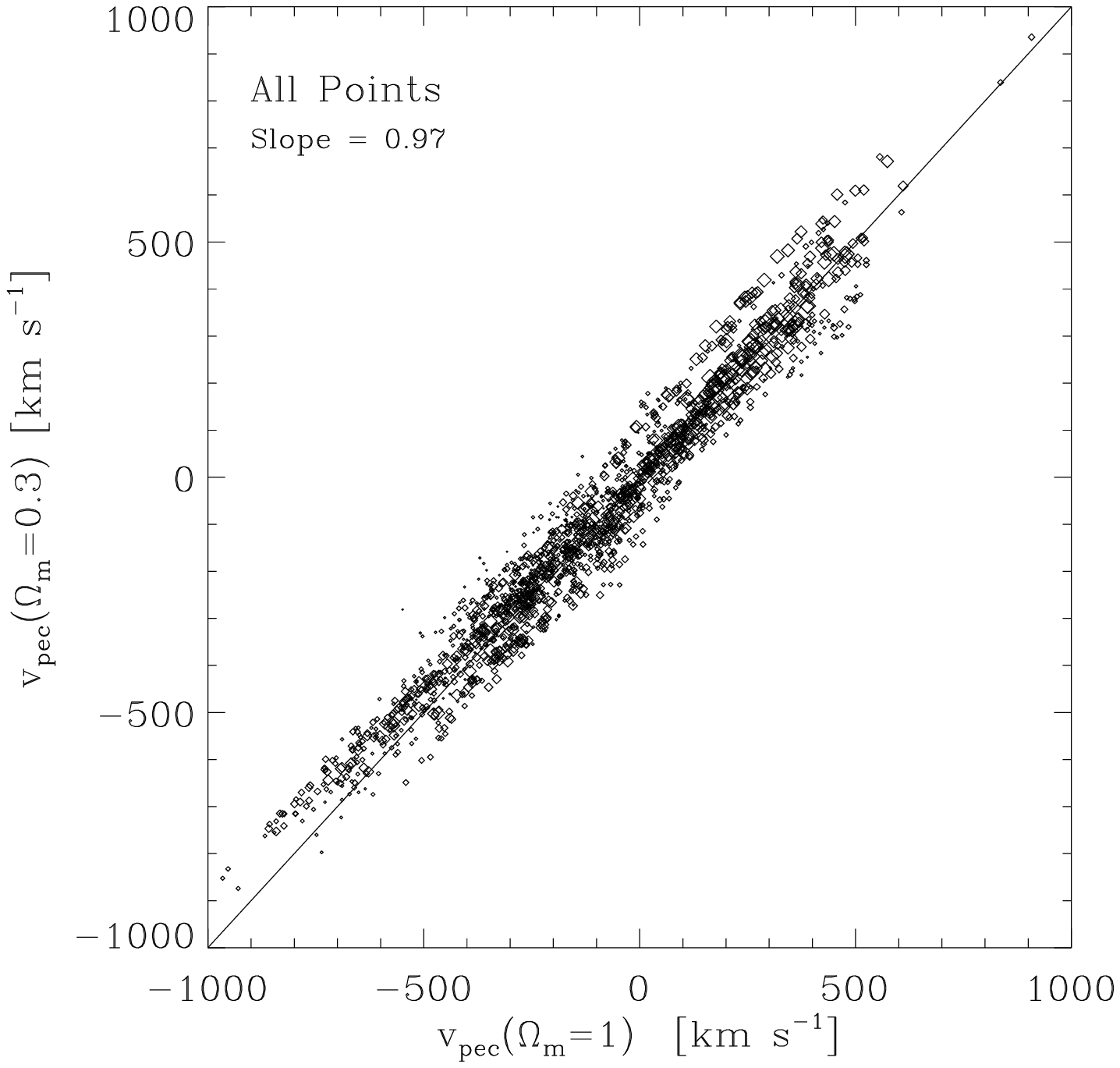}}
\end{picture}
\vskip -3 truecm
\caption{\capt The top left panel compares the 
G12-smoothed peculiar radial velocities 
reconstructed from the SFI catalog  {\it versus} 
those reconstructed from the SEcat catalog, both measured at 
the location of the SFI  data.
The solid line represents the best linear fit to the scatter plot. 
The top-right panel compares ENEAR and SEcat reconstructed velocities
at the positions of the  ENEAR galaxies.
The two bottom panels refer to the SEcat catalog only and compare 
velocities reconstructed assuming a $\Lambda$CDM model (Y axis)
with those reconstructed from the standard CDM model (X axis).
Peculiar velocities shown in the bottom-left panel refer to points
with estimated error less than $100/kms$. All points are considered 
in the bottom-right panel.
In all plots the size of the symbols is inversely
proportional to their errors.
}
\label{fig:velocity_compare}
\end{figure*}

Before proceeding further in analysing the SEcat catalog, it is useful
to compare the reconstruction we obtain from the SEcat peculiar
velocities with those obtained from the SFI and ENEAR separately in
order to support our claim regarding the consistency the latter
two. Figures \ref{fig:density_compare}, \ref{fig:enear_vs_sfi}, and
\ref{fig:velocity_compare} show the similarity between the density and
velocity reconstruction of the three catalogs.  We also note that, as
demonstrated in the same figures, changing the power spectrum prior
has a very small effect on the results.

The reconstruction of the PSC$z$ model density and velocity fields,
performed according to Branchini \etal (1999), requires an input value
for $\beta$; here we use $\beta=0.5$. However, the dependence of the
reconstructed density field on the input $\beta$ is very weak while
the model peculiar velocity roughly scales linearly with $\beta$. This
dependence results in a systematic error of about 2\% on the final
result. Model PSC$z$ velocities were obtained at the reconstructed
positions of PSC$z$ galaxies in real space. Both masses and velocities
were then smoothed with a G12 filter to obtain the PSC$z$ density and
velocity fields on a regular grid with a mesh size of $2.5 \hmpc$
Finally, G12-smoothed PSC$z$ velocities have been interpolated at the
positions of SEcat galaxies. Here we compare the SEcat fields with
those of the PSC$z$, all smoothed with a G12 filter. The errors on the
model density and velocity fields are much smaller than the SEcat ones
(see Branchini \etal, 1999) and will be ignored in the following
comparisons.

\begin{figure*}
\setlength{\unitlength}{1cm} \centering
\begin{picture}(18.,18.)
\put(-2.5, -11.){\includegraphics{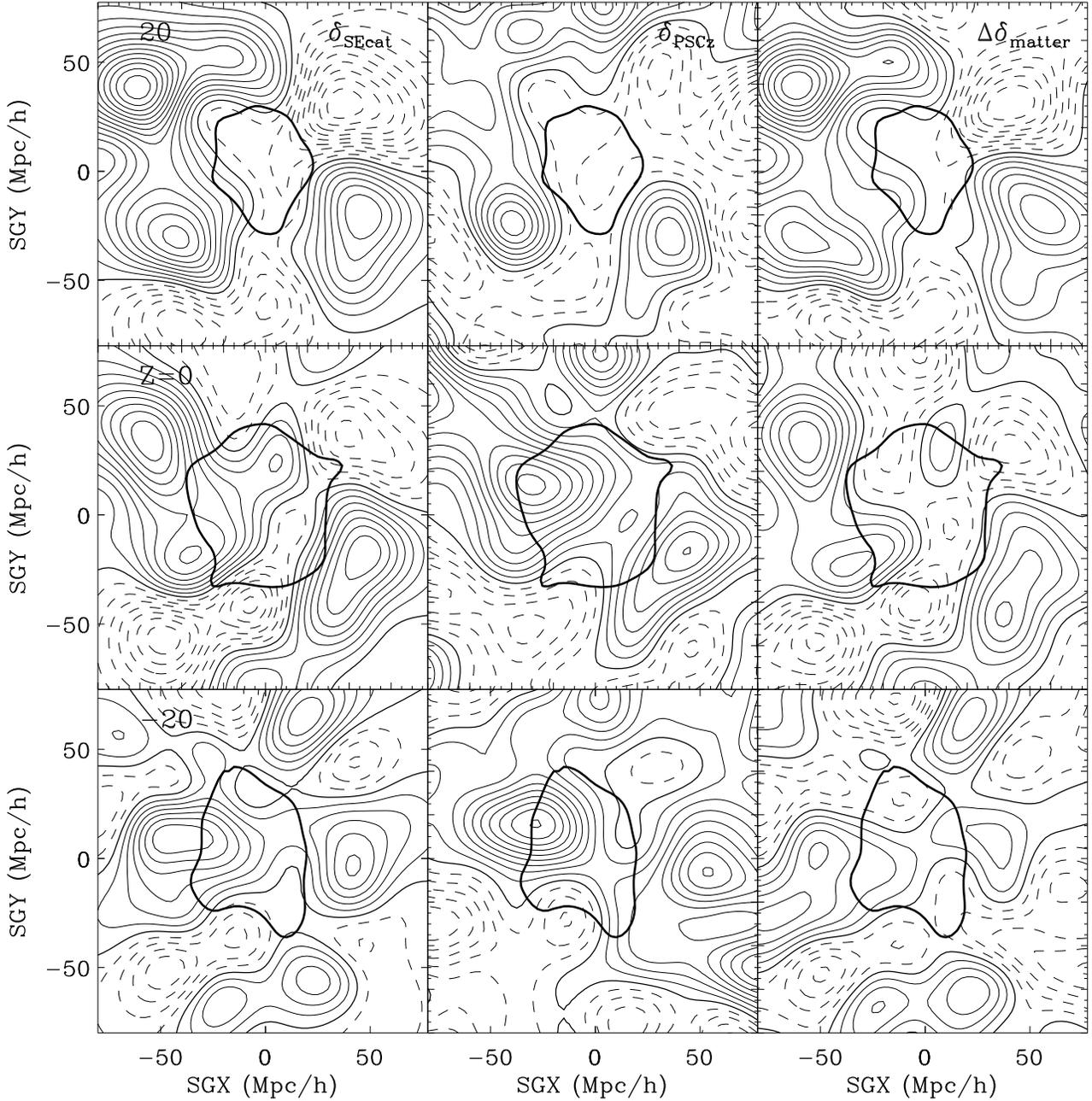}}
\end{picture}
\caption{\capt The panels on the left show the maps of the
G12-smoothed, density fluctuations UMV-reconstructed 
from the SEcat catalog. The central panels show the same field
reconstructed from the distribution of  PSC$z$ galaxies.
The maps of density residuals, computed for $\beta=0.57$  $\Delta\delta=
\delta_{SEcat}-\delta_{PSCz}$ are shown on the right-hand panels.
The central maps show the density fields on the Supergalactic planes.
The maps on the upper and lower panels show the density fields
at Supergalactic Z$=\pm 20 ^{-1}\; Mpc$, respectively.
The very thick contour marks the boundaries of
the volume within which the estimated reconstructed error is $\le
0.2$; the volume used for comparisons.  }
\label{fig:umv_g12}
\end{figure*}    
\begin{figure}
\setlength{\unitlength}{1cm} \centering
\begin{picture}(9.,18.)
\put(-2.2, 7.5){\includegraphics{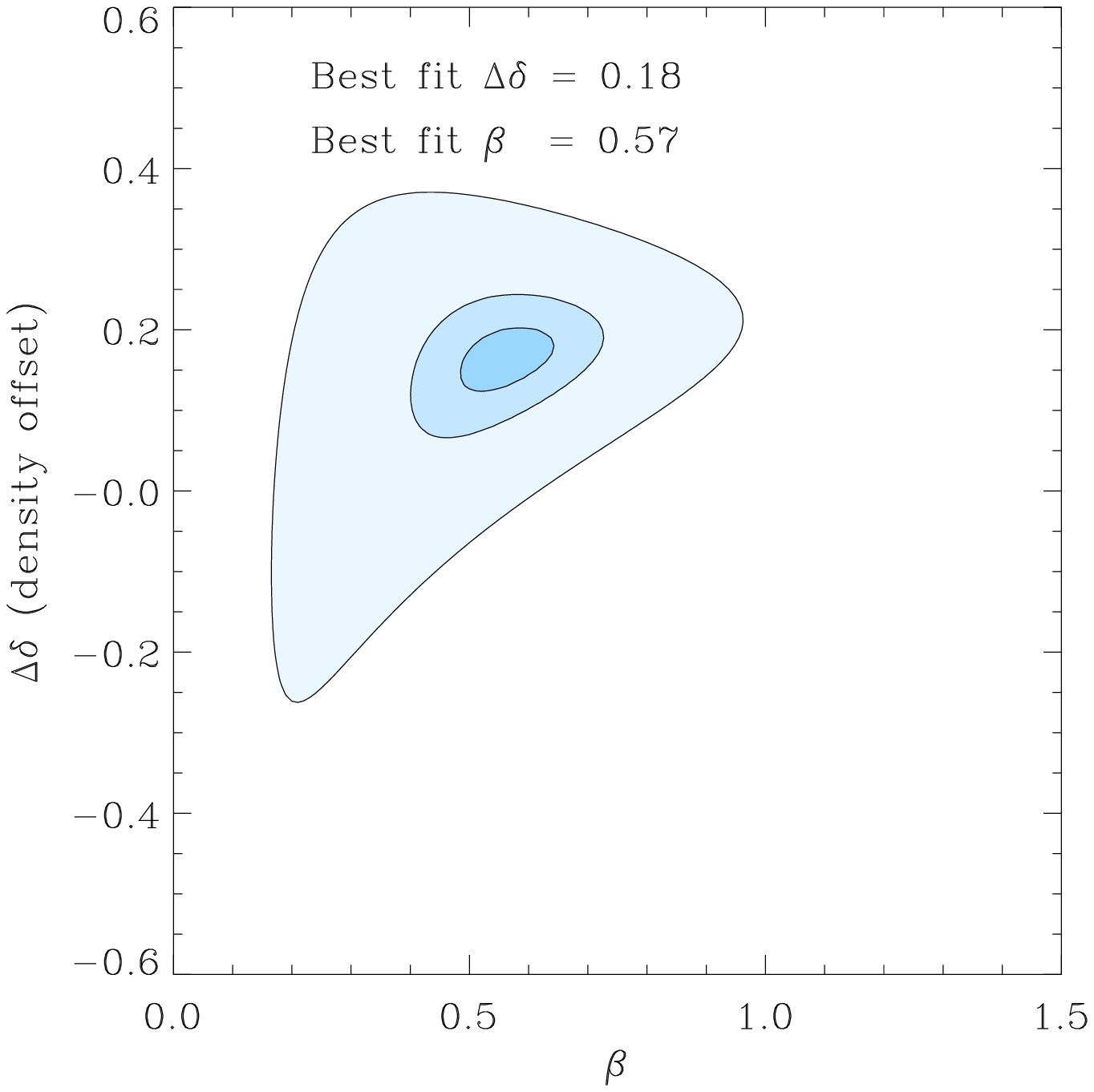}}
\put(-2.2,-1.7){\includegraphics{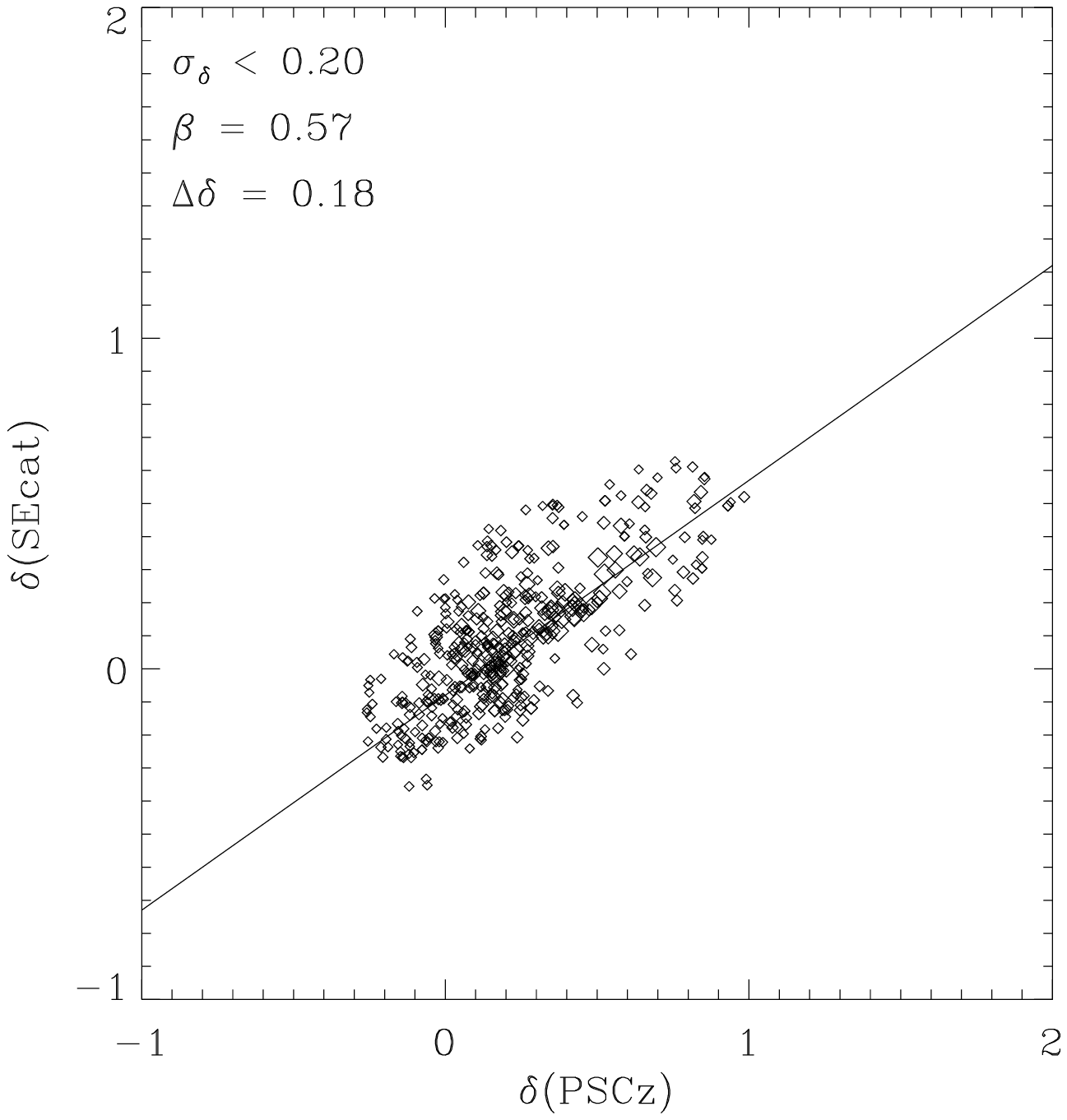}}
\end{picture}
\caption{\capt The upper panel shows the 1, 2 and 3 $\sigma$ likelihood
contours from the $\delta-\delta$ comparison
in the $\beta - \Delta\delta$ plane. 
The scatter plot in the bottom panel compares 
the G12-smoothed UMV-reconstructed SEcat density field
to the PSC$z$ density fluctuations, both measured at 
grid points 1/10 randomly selected from those with 
 reconstruction errors $\sigma_\delta < 0.2$ 
The size of the symbols is inversely
proportional to their errors. }
\label{fig:dd}
\end{figure}    

\subsection{The density-density comparison}
\label{sec:dd}
The left panels of Figure~\ref{fig:umv_g12} show the G12 smoothed UMV
reconstructed density field map in three planes at different 
Supergalactic Z (the central plane refers to Z=0, i.e. the Supergalactic
plane) obtained
from the SEcat peculiar velocities within a box of  $160\hmpc$ aside,
centered around the Local Group position. The main features of our
local universe are easily identified 
in the UMV map on the Supergalactic plane , including the
Great Attractor on the left and the Perseus-Pisces supercluster in the
lower right. There is also a hint of the Coma cluster, which lies just
outside the sample, in the upper part on the map.  Similar features
also characterize the PSC$z$ density map shown on the right-hand panel
of Figure~\ref{fig:umv_g12}. This map, obtained by Branchini \etal
(1999), has the same smoothing (G12) and shows the same region of
the universe.
 
A quantitative density-density comparison is carried out using the
following $\chi^2$ statistic:
\begin{equation}
\chi^2 = {1 \over N}\sum_{\sigma_\delta \le 0.2} {\left[\delta_i({\rm
SEcat}) - \beta (\delta_i({\rm PSC}z) + \Delta\delta)\right]^2\over 
\sigma_\delta^2},
\label{eq:chi2}
\end{equation}
where $\delta_i(SEcat)$ and $\delta_i({\rm PSC}z)$ are the SEcat and
PSC$z$ densities respectively, $\Delta\delta$ is the offset in the
mean density and $\sigma_\delta$ are the density reconstruction error
at each point as estimated from Monte Carlo realizations of mock-SEcat
catalogs. The best fit $\beta$ and $\Delta\delta$ parameters are 
those that minimize Eq.~\ref{eq:chi2}.  Here again the sum is over 
gridpoints  randomly sampled at a rate of 1/10 from those that have
errors less than 0.2. 
As in the mock catalogs analysis, here ${\cal N}_{dof} \approx 20$.

In previous density-density comparisons (\eg, Sigad \etal, 1998) the
authors chose to minimize the $\chi^2$ with respect to 
$1/\beta$ instead of $\beta$ directly. Since the main source of
errors in our analysis are the uncertainties in the measured 
galaxy velocities, adopting a direct or inverse $\beta$ minimization
of the $\chi^2$ statistic does not make much difference.
Here we choose to minimize with respect to $\beta$.

Each point in the scatter plot displayed in the lower panel of
Figure~\ref{fig:dd} shows the comparison between the SEcat and PSC$z$
overdensities, measured at the same locations. The comparison is
restricted to the 1/10 randomly chosen points with
$\sigma_\delta < 0.2 $. The slope of the solid line gives
$\beta=0.57$.

A zero-point offset $\Delta \delta =0.18$ is also detected.
We interpret it as a
mismatch in the average density in the two samples. The mismatch in
the mean fields is caused by the PSC$z$ density field which was found
to be systematically larger than the IRAS 1.2Jy density field within a
$60\hmpc$ sphere (Teodoro \etal, 2000). This mismatch arises due to the
incompleteness of the PSC$z$ catalog at low fluxes.

The upper panel in Figure~\ref{fig:dd} shows the 1, 2 and 3-$\sigma$
likelihood contours in the $\beta$--$\Delta\delta$ plane obtained from
Eq.~\ref{eq:chi2}. The marginalization of this distribution with respect
to $\Delta\delta$ gives the error estimate on the value of $\beta$
which is of the order of 0.08 ($\approx 15\%$). Adding to this error
the error estimated from the distribution of bias shown in
Figure~\ref{fig:betafrommock} one obtains
$\beta=0.57_{-0.13}^{+0.11}$.

\begin{figure}
\setlength{\unitlength}{1cm} \centering
\begin{picture}(9.,9.)
\put(-2, -1.5){\includegraphics{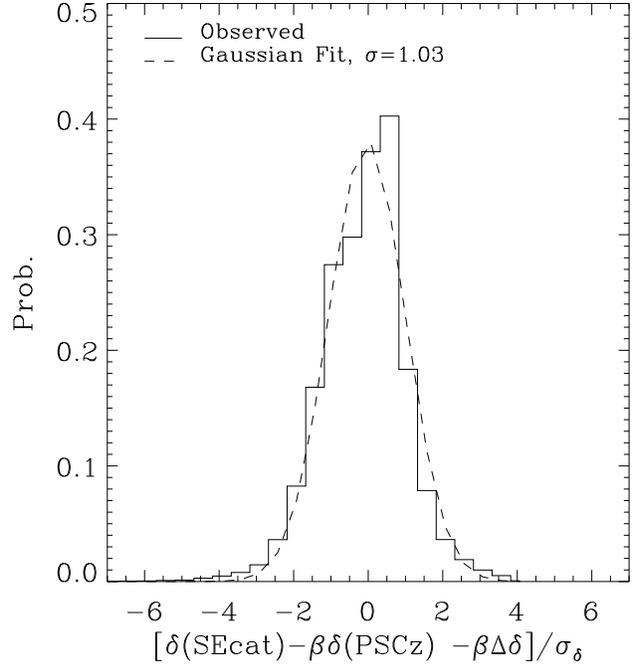}}
\end{picture}
\caption{\capt 
The histogram represents the distribution of the normalized residuals
($\xi$). The dashed line shows its best Gaussian fit with a rms
value of $\sigma=1.03$, as indicated in the plot.}
\label{fig:residual_dd}
\end{figure}

To estimate the goodness-of-fit of the parameters obtained from the
$\chi^2$ analysis, we calculate the distribution of the residuals,
\begin{equation}
\xi={\delta({\rm SEcat}) - \beta_{ML}\bigl(\delta({\rm PSC}z) -\Delta\delta\bigr)\over \sigma_\delta}, 
\label{eq:resdiual}
\end{equation}
where $\beta_{ML}$ is the best fit $\beta$ parameter and $\sigma_\delta$ is
the error on the UMV estimated density field. If the model
correctly describes the data, this distribution should be Gaussian
with a rms of unity. The histogram of $\xi$ is shown in
Figure~\ref{fig:residual_dd} along with the best fitting Gaussian
distribution (dashed line), whose rms is almost unity.

To check the robustness of the result we have repeated the
density-density comparison by drawing three additional samples defined
at different error thresholds and re-estimating the values of
$\beta$. Adopting the thresholds $\sigma_\delta=0.3, 0.4$ and $0.5$ we
obtained $\beta = 0.54, 0.53$ and $0.52$, respectively.

\subsection{The velocity-velocity comparison}
\label{sec:vv}

\begin{figure}
\setlength{\unitlength}{1cm} \centering
\begin{picture}(9,17.5)
\put(-3, -1.){\includegraphics{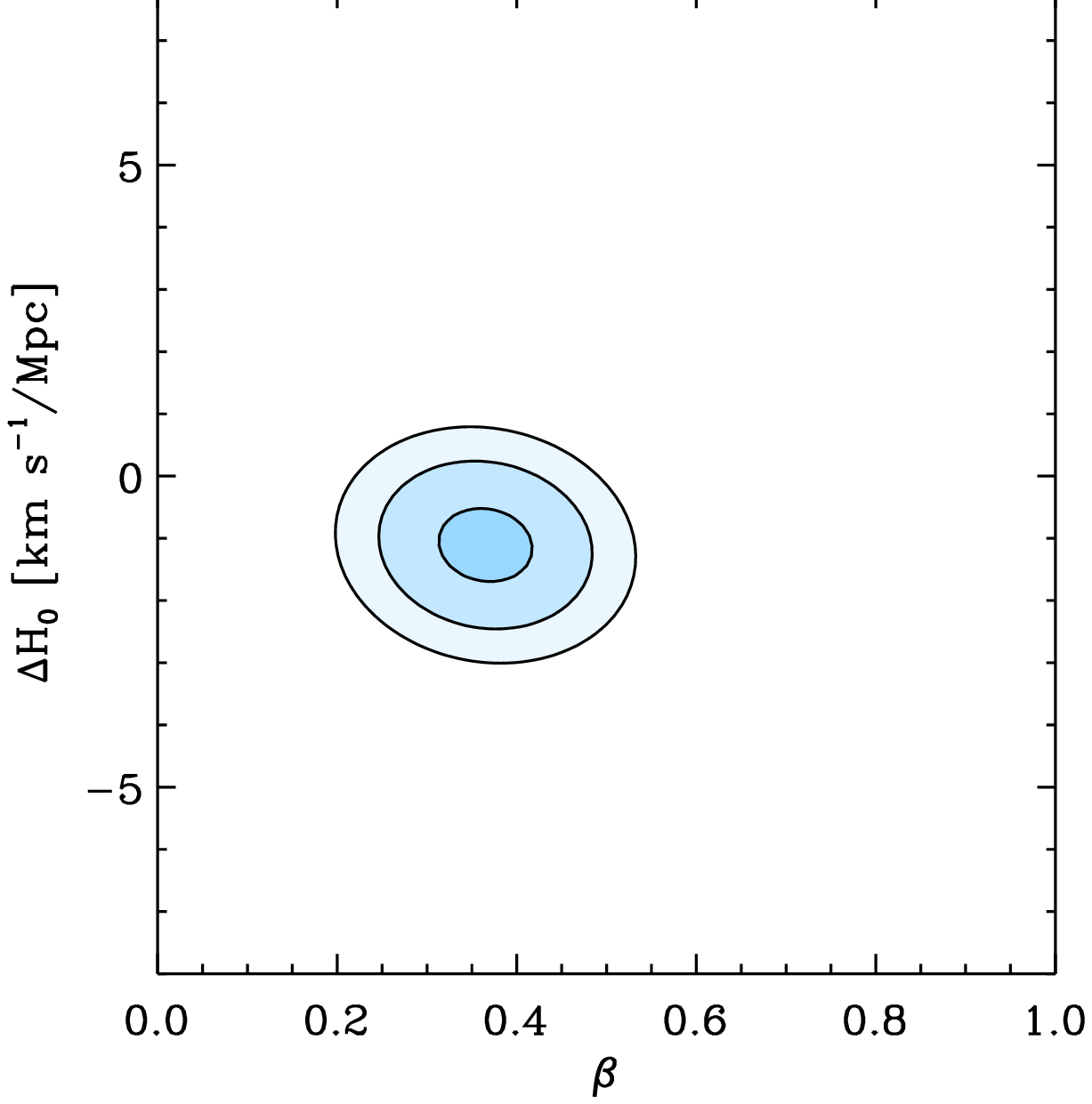}}
\put(-3, -1.){\includegraphics{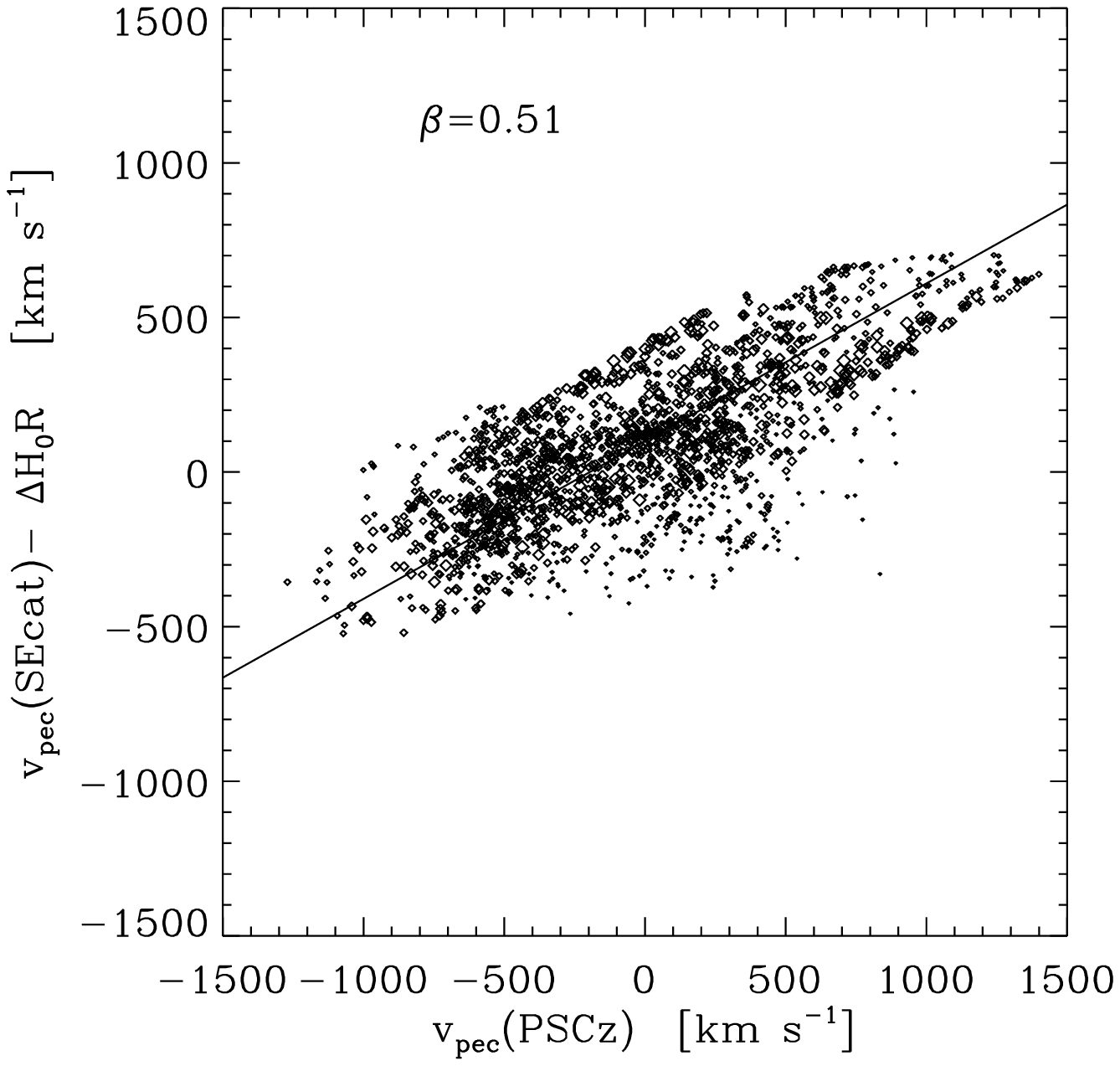}}
\end{picture}
\caption{\capt Upper Panel: The 1,2, and 3 $\sigma$ likelihood contours
in the $\beta-\Delta H_{\circ}$ plane
from the SEcat {\it vs.} PSC$z$ velocity-velocity comparison.
In the bottom panel the G12-smoothed, reconstructed SEcat and  PSC$z$  
are compared  at the locations of the SEcat data points.
The size of the symbols is inversely proportional to the reconstruction errors}
\label{fig:vv}
\end{figure}

The UMV-velocity reconstruction procedure is different from the one
used for the density. SEcat radial velocities are calculated at the
location of the data points from the reconstructed 3D velcoty which
has been homogeneously smoothed with a G12 filter through the UMV
operation.  Here we compare these velocities to the G12-smoothed model
PSC$z$ velocity sampled at the same locations.  Therefore, unlike for
the density-density case, the number of points used in the
velocity-velocity comparison is determined directly by the number of
points in the SEcat catalog. This comparison is carried out using the
following $\chi^2$ statistic
\begin{equation}
\chi^2 = {1 \over N}\sum_{Data Points} {\left(u_i({\rm SEcat}) - \beta u_i({\rm PSC}z) -
\Delta H_{\circ}r_i\right)^2\over \sigma_v^2},
\label{eq:chi2_vv}
\end{equation}
where $u$ denotes the G12-smoothed radial velocities, $\Delta
H_{\circ}$ is a local perturbation to the Hubble constant, $r_i$ is
the radial distance of the point $i$ and $\sigma_v$ is the error in
the UMV reconstruction.  The resulting velocity-velocity scatter-plot
is shown in the lower panel of Figure~\ref{fig:vv} along with the best
fitting line.  Here the slope of the line constitutes an estimate of
$\beta=0.51$. The zero-point mismatch, $\Delta H_{\circ}$ represents a
spurious ``breathing-mode'' which is to be expected given the average
density mismatch found in the density-density comparison.  A
perturbation of $\Delta H_{\circ}=1.5$\kms/$Mpc$ was found and its
associated spurious radial motion, $\Delta H_{\circ} r$, was
subtracted from the PSC$z$ peculiar velocities shown in the lower
panel of Figure~\ref{fig:vv}.  The two zero-points are found here to
be consistent with the prediction of the linear theory relation:
\begin{equation}
\Delta H_{\circ} = -{\Omega_m^{0.6} \over 3} \Delta\delta(< r) H_{\circ},
\label{dh}
\end{equation} 
where $\Delta\delta(< r)$ is the mean-density mismatch within a radius
$r$.

The upper panel of Figure~\ref{fig:vv} shows the likelihood contours
in the $\beta-\Delta H_{\circ}$ plane obtained by calculating the
$\chi^2$ distribution given in Eq.~\ref{eq:chi2_vv}.  From the
iso-probability contours of 1,2 and 3-$\sigma$ levels shown in the
upper panel of Figure~\ref{fig:vv} we obtain that $\beta=0.51\pm
0.06$, fully consistent with the estimate of $\beta$ from the
density-density comparison. The uncertainty here is a combination of
the errors estimated from Figures~\ref{fig:betafrommock}
and~\ref{fig:vv}.

\begin{figure}
\setlength{\unitlength}{1cm} \centering
\begin{picture}(9.,9.)
\put(-2.,-1.5){\includegraphics{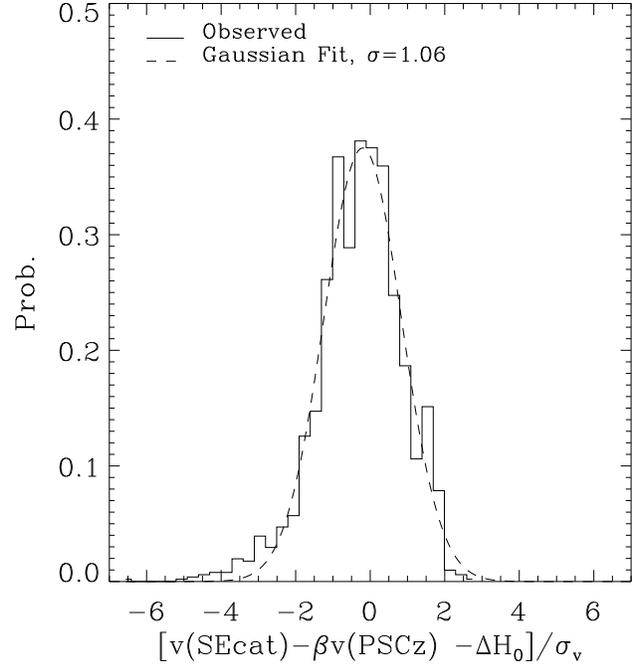}}
\end{picture}
\caption{\capt The distribution of the residuals from the velocity
velocity comparison. The dashed line shows a Gaussian fit with $\sigma=1.06$. }
\label{fig:residual_vv}
\end{figure}    

The analysis of the velocity-velocity residuals is performed similarly
to those of the density-density residuals. The resulting distribution
of the residuals is shown in Figure~\ref{fig:residual_vv} as a
histogram. Here again the rms value found for the best fit Gaussian
distribution (dashed line) is very close to unity indicating the
adequacy of the PSC$z$ velocity model with $\beta=0.51$. The slight
excess in the positive tail of the histogram is due to presence of few
outliers,easily identified below the best fitting line in the lower
panel of Figure~\ref{fig:vv}. The size of those points clearly indicates
that they have very large errors, which would exclude them from any
reasonable noise data-cut analysis. Moreover, their number is
very small and therefore their influence on the final result is
negligible.

\section{Summary and Discussion}
\label{sec:discussion}

In this paper we have applied the new UMV estimator to recover the
density and velocity fields in the local universe from the SEcat
catalog of galaxy peculiar velocities. In order to obtain the so
called $\beta$ parameter, these fields have been compared with 
those modeled from the spatial distribution of IRAS PSC$z$ galaxies
assuming linear theory and biasing. Previous estimates of $\beta$ from
density-density comparisons, mainly based on the POTENT algorithm
(Bertschinger \& Dekel 1989, Dekel \etal, 1990), have yielded a large
value ($\beta\approx 1$ \cf, Sigad \etal, 1998), inconsistent with
the smaller values ($\approx 0.5$) independently obtained from all recent 
velocity-velocity VELMOD (Willick \etal, 1996, Willick \&
Strauss 1998, and Branchini \etal, 2001) and ITF (da Costa \etal, 1998
and Nusser \etal, 2000) comparisons.

For the first time the UMV method provides a common methodological
framework in which to perform velocity-velocity and density-density
comparisons.  The velocity-velocity comparison yields a value of
$\beta$ consistent with that measured in the VELMOD and ITF
analyses. However, the value of the same parameter obtained from our
density-density comparison is significantly smaller than those
obtained from the POTENT analyses (\cf, Sigad \etal, 1998).  The
$\beta$ parameters from both $v-v$ and $\delta-\delta$ comparisons
presented here are in agreement, yielding a $\beta \approx 0.55$ with
an estimated error of the order of $0.1$.

In contrast with the POTENT algorithm, the new UMV method reconstructs
the density field from peculiar velocities while taking into account
their underlying correlation properties. The regularization aspect of
the UMV estimator significantly improves the stability of the
inversion, which is especially important given the low signal-to-noise
ratio of peculiar velocity data.  The regularization obtained by this
method is very similar to the one provided by the Wiener filter method
(Zaroubi \etal, 1999). However, the UMV has been designed to provide an
unbiased estimator of the underlying signal, a property that the
Wiener filtering method lacks. These two aspects make the UMV
estimator a very useful tool for reconstruction from peculiar velocity
data.

In our error analysis we have shown that for the best fit value of
$\beta$ the density and velocity residuals are normally
distributed. This indicates that the PSC$z$ density and velocity
fields constitute an adequate model for those reconstructed with the
UMV estimator. The fields only differ by a monopole term,
corresponding to a mismatch in the the mean density within $60 \hmpc$
which is caused by the known incompleteness of the PSC$z$ catalog at
faint fluxes (Teodoro \etal, 2000).  This also implies that the effect
of the nonlinear dynamics and amount of nonlinear and stochastic
biasing on the scales involved in our analysis is negligible relative
to the measured peculiar velocity errors.

The results presented in this paper are quite encouraging since for
the first time the two ways of estimating the value of the $\beta$
parameter give a consistent result. Our results also suggest that the
UMV estimator is a promising tool for the problem of reconstructing
the dynamical fields from the observed radial peculiar velocities and
therefore could be applied to other datasets. In particular, to
reconstruct the large scale structure from the incoming large and
uniform surveys that will provide both the spatial distribution and
peculiar velocities of extra-galactic objects \eg, SDSS and large
cluster surveys with kinematic Sunyaev-Zel'dovich measurements.

Our present density-density comparison results are in marked contrast
to those obtained by earlier work, including those from the recent
analysis of the Mark~III catalog using the POTENT method (\eg, Sigad
\etal, 1998), and raises the question on the origin of this
discrepancy. The Mark III catalog, as shown for example by Davis
\etal\ (1996) and more recently by Courteau \etal\ (2000), suffers
from systematic calibration errors that would cause a systematic error
in the estimation of $\beta$.  However, these errors are not expected
to overestimate the value of $\beta$ by more than a factor of two.  An
application of the UMV method to the Mark III catalog shows that the
obtained values of $\beta$ are somewhat higher than those obtained
from the SEcat catalogs by $0.1-0.2$. Moreover, the $v-v$-like VELMOD
analysis yield consistent values of $\beta$ when applied to Mark III
and SFI datasets (Willick \etal, 1997b, Willick and Strauss 1998,
Branchini \etal, 2001).  Based on these arguments, one could speculate
that the most likely explanation to the inconsistent results is a
conspiracy of both the systematics errors in Mark III and some
noise-driven inversion instability in the POTENT reconstructions.

\section*{Acknowledgments}

The authors would like to thank the referee, Michael Strauss, for his
very useful comments and suggestions, and Anthony J. Banday for
critical reading of the manuscript. LdC and SZ would also like to
thank the entire ENEAR team. YH and SZ thanks the Universit\'a di Roma
Tre. EB and YH thank the Max Planck Institut f\"{u}r Astrophysik for
their hospitality while part of this work was done. YH has been
supported in part by the Israel Science Foundation (103/98).

\end{document}